\documentclass{aa}
\usepackage{graphicx}
\newcommand{\logg}{\log\,g}
\newcommand{\teff}{T_{\rm eff}}
\def\Rh{\rule{5.0pt}{0.0pt}}
\def\Rv{\rule[-0.0in]{0.0pt}{10.0pt}}
\def\Rb{\rule[-5.0pt]{0.0pt}{5.0pt}}
\def\Rr{\rule{15.0pt}{0.0pt}} 
\begin{document}

\title{Line identification in high-resolution,
near-infrared CRIRES spectra of chemically peculiar and Herbig Ae stars\thanks
{Based on observations obtained at the European Southern Observatory (ESO programme 087.C-0124(A)).}}

\author{
S.~Hubrig\inst{1}
\and
F.~Castelli\inst{2}
\and
J.~F.~Gonz\'alez\inst{3}
\and
V.~G.~Elkin\inst{4}
\and
G.~Mathys\inst{5}
\and
C.~R.~Cowley\inst{6}
\and
B.~Wolff\inst{7}
\and
M.~Sch\"oller\inst{7}
}

\institute{
Leibniz-Institut f\"ur Astrophysik Potsdam (AIP), An der Sternwarte 16, 14482 Potsdam, Germany
\and
Istituto Nazionale di Astrofisica, Osservatorio Astronomico di Trieste, via Tiepolo 11, 34143 Trieste, Italy 
\and
Instituto de Ciencias Astronomicas, de la Tierra, y del Espacio (ICATE), 5400 San Juan, Argentina
\and
Jeremiah Horrocks Institute, University of Central Lancashire, Preston PR1 2HE
\and
European Southern Observatory, Casilla 19001, Santiago, Chile 
\and
Department of Astronomy, University of Michigan, Ann Arbor, MI 48109-1042, USA
\and
European Southern Observatory, Karl-Schwarzschild-Str.~2, 85748 Garching bei M\"unchen, Germany
}

\abstract
 % context heading (optional)
{Contrary to the late-type stars, our knowledge of atomic transitions in intermediate-mass stars
is still very poor. The recent availability of ESO's high-resolution spectrograph CRIRES offers 
now the opportunity
to study numerous spectral features in the near-IR in intermediate-mass main-sequence and 
pre-main-sequence stars.}  
{
The aim of the study is to explore the diagnostic potential of near-IR spectral
regions. We carry out the first line identification in a few spectral regions for the two strongly
magnetic Ap stars $\gamma$\,Equ and HD\,154708, and their potential precursors two pre-main sequence Herbig Ae/Be 
stars HD\,101412 and 51\,Oph.}
{
High-resolution CRIRES spectra were obtained in three spectral regions, two regions around 1\,$\mu$m and 
one region around 1.57\,$\mu$m containing magnetically sensitive \ion{Fe}{i} lines. To study the spectral
line variability in the Herbig Ae star HD\,101412, the observations were collected on six different
rotation phases.
All currently available atomic line lists were involved to properly identify the detected spectral features.
}
{
The largest number of near-IR spectral features was detected and identified in the  
well- studied 
magnetic Ap star $\gamma$\,Equ. Nearly 30\% of the spectral lines
in the Ap star HD\,154708 with one of the strongest magnetic fields known among the Ap stars of 
the order
of 25\,kG, remain unidentified due to a lack of atomic data. Only very few lines belonging to the rare earth 
element group have been identified in both Ap stars. A number of spectral lines
including the \ion{Ce}{iii} and \ion{Dy}{ii} lines appear magnetically split due to the presence of a 
strong magnetic field in their
atmospheres.
The content of the spectra of the Herbig Ae/Be stars HD\,101412 and 51\,Oph is rather unexciting.
Variable behaviour of lines of the elements He, N, Mg, Si, and Fe over the 
rotation period in the spectra of HD\,101412 confirm our previous finding of variability in 
the optical region.
Due to the very fast rotation of 51\,Oph, only a few spectral lines have been identified with certainty.
}
{}


\titlerunning{Line identification in the near-IR}
\authorrunning{S.\ Hubrig et al.}
\maketitle

%________________________________________________________________

\section{Introduction}

Owing to the extreme richness of UV and optical spectra, the upper main-sequence
chemically peculiar (CP) stars have been intensively 
studied at the highest possible spectral resolution over the last decades. Among the different groups 
of CP stars,
especially classical Ap stars with strong magnetic fields of $\sim$~kG order and exhibiting strong overabundances
of iron-peak and rare earth elements are of particular
interest. These stars present a natural laboratory to study the element enrichment of stellar atmospheres
due to the operation of various competing physical effects (such as, e.g., microscopic diffusion
of trace atomic species) in the presence of strong magnetic fields. Due to their generally slow rotation
it is also possible to study the isotopic and hyperfine structure of certain elements and their interaction
with the magnetic field. 

The progenitors of main-sequence stars of intermediate mass are pre-main 
sequence Herbig Ae/Be stars.
Recent spectropolarimetric observations of a few Herbig Ae/Be stars indicate that
magnetic fields are important ingredients of the intermediate-mass star formation process. 
As an example, the sharp-lined young Herbig Ae star HD\,101412 with a strong surface magnetic field of the 
order of a few kG has become over the past 
few years one of the most studied targets among the Herbig Ae/Be stars using optical 
and polarimetric spectra.  

The recent availability of ESO's high-resolution CRyogenic Infra-Red Echelle
Spectrograph (CRIRES) installed at the ANTU telescope on Cerro Paranal offers now the opportunity to acquire
much better knowledge of spectral features in intermediate-mass stars in the near-IR.
A study of 
atomic transitions in this wavelength region for classical magnetic Ap stars is of great interest 
as the Zeeman 
splitting of spectral lines in the presence
of a magnetic field is proportional to $\lambda^2$. Knowledge of magnetically sensitive 
atomic transitions in the spectra of intermediate-mass stars will allow one to achieve the 
detection of 
weaker stellar magnetic fields in future studies than is feasible in the optical wavelength region. 

Regarding the Herbig Ae/Be stars, a study of their chemical anomalies and a search
for the presence of exotic elements in their atmospheres is important to 
understand the origin of chemically peculiar Ap/Bp stars (e.g., Hubrig et al.\ \cite{Hubrig2000,Hubrig2007,
Hubrig2009a}).
Most near-IR studies of Herbig Ae/Be stars
were devoted to molecular spectroscopy to probe the conditions and physical processes in protoplanetary
disks. Among the known magnetic Herbig Ae/Be stars, the slow rotation and strong magnetic 
field of the Herbig Ae star HD\,101412 make this star a prime candidate for a line identification study
in the near-IR. 

The aim of the present study is to explore the diagnostic potential of a few  near-IR spectral 
regions in A and B-type stars.
 We carried out in a few wavelength regions the first line identification for two strongly
magnetic Ap stars, $\gamma$\,Equ and HD\,154708, and the magnetic Herbig Ae/Be star HD\,101412.
In addition, we present CRIRES observations of one of the fastest rotating Herbig Be stars (51\,Oph)
and discuss the diagnostic potential of CRIRES observations for rapidly rotating stars.

\section{Observations}

The observations were obtained in service mode between April and
June 2011 with the near-infrared (NIR) spectrograph CRIRES using a slit width of 0.2\arcsec{}.
Observations
followed an AB on-slit nodding scheme, collecting two to
eight spectra with individual exposure times of 10 to 300\,s, depending
on the brightness of the individual target.
Routines of the ESO CRIRES pipeline software have been used for data reduction. Raw observations 
are flat-fielded, corrected for non-linear detector response, and for bad pixels before frames from
different nodding positions are shifted and combined. This procedure results in a small degradation 
of the spectral resolution because of optical distortions of the instrument along the slit. Therefore, 
we reach a resolution of 94,600 (instead of the nominal 100,000), which was determined from weak telluric
lines.
From the combined images, spectra are extracted using an optimum extraction algorithm. Wavelength calibration 
is determined from daytime ThAr arc lamp exposures.

\begin{table}
\caption{
Logbook of CRIRES observations.
}
\label{tab:log}
\centering
\begin{tabular}{lcccc}
\hline
\hline
\multicolumn{1}{l}{Object} &
\multicolumn{1}{c}{MJD} &
\multicolumn{1}{c}{Phase} &
\multicolumn{1}{c}{Central} &
\multicolumn{1}{c}{S/N} \\
 & & &
\multicolumn{1}{c}{wavelength} \\
 & & &
\multicolumn{1}{c}{[nm]} \\
\hline
HD\,101412 & 55654.10 & 0.89 &1082.7 &   142--203 \\ 
           & 55654.10 & 0.89 &1100.5 &   104--133 \\ 
           & 55654.10 & 0.89 &1574.4 &   441--558 \\ 
           & 55656.10 & 0.94 & 1082.7 &  114--174  \\
           & 55656.10 & 0.94 & 1100.5 &  110--123  \\
           & 55656.10 & 0.94 & 1574.4 &  486--580  \\
           & 55662.09 & 0.08 & 1082.7 &  153--187  \\
           & 55662.09 & 0.08 & 1100.5 &  189--234  \\
           & 55662.09 & 0.08 & 1574.4 &  560--564  \\
           & 55667.21 & 0.21 & 1082.7 &  274--330  \\
           & 55667.21 & 0.21 & 1100.5 &  212--262  \\
           & 55667.21 & 0.21 & 1574.4 &  569--690  \\
           & 55681.07 & 0.53 & 1082.7 &  262--306  \\
           & 55681.07 & 0.53 & 1100.5 &  212--299  \\
           & 55681.07 & 0.53 & 1574.4 &  563--655  \\
           & 55683.07 & 0.58 & 1082.7 &  231--298  \\
           & 55683.07 & 0.58 & 1100.5 &  191--279  \\
           & 55683.07 & 0.58 & 1574.4 &  522--553  \\
HD\,154708 & 55700.27 & 0.87 & 1082.7 &  171--230 \\
           & 55700.27 & 0.87 & 1100.5 &  157--215 \\
           & 55700.27 & 0.87 & 1574.4 &  372--415 \\
51\,Oph      & 55704.30 &  -- & 1082.7 &  407--577 \\
             & 55704.30 &  -- & 1100.5 &  335--381 \\
             & 55704.30 &  -- & 1574.4 &  618--773 \\
$\gamma$\,Equ & 55728.31 &  -- & 1082.7 &  270--378 \\
             & 55728.31 &  -- & 1100.5 &  385--393 \\
             & 55728.31 &  -- & 1574.4 &  1342--953 \\
\hline
\end{tabular}
\end{table}

The selected wavelength regions contain the region around the \ion{He}{i} 1083.0\,nm line 
(region 1065.5--1091.4\,nm),
the region around the Pa$\gamma$ line (1084.3--1108.5\,nm), and the region with strongly magnetically 
sensitive \ion{Fe}{i} lines (1550.2--1586.6\,nm).
CRIRES makes use of four Aladdin detectors located in the focal plane of the spectrograph.
The detectors 1 and 4 are less useful due to contamination by the adjacent orders. This contamination
is worse at shorter wavelengths. As a result, the wavelength calibration on these
detectors in the spectral region around 1\,$\mu$m
is not reliable and makes the identification work difficult.

Table~\ref{tab:log} summarizes information on the observed targets,
including the observed wavelength regions and signal-to-noise ratios for the best detectors 2 and 3.
The phases for HD\,101412 were calculated according to the ephemerides presented in the work of
Hubrig et al.\ (\cite{Hubrig2011}) using magnetic and photometric data:

\begin{equation}
\left<B\&I\right>^{\rm max} = {\rm MJD}52797.4 \pm 0.8 + 42.076 \pm 0.017 E.
\end{equation}

For HD\,154708 the phase at the observational epoch with CRIRES was calculated 
according to the ephemerides presented in the work of
Hubrig et al.\ (\cite{Hubrig2009b}) using magnetic data:

\begin{equation}
\left<B_{\rm z}\right>^{\rm max} = {\rm MJD}54257.26 \pm 0.03 + 5.367 \pm 0.020 E
\end{equation}

\begin{figure}
\centering
\includegraphics[width=0.45\textwidth, height=7cm]{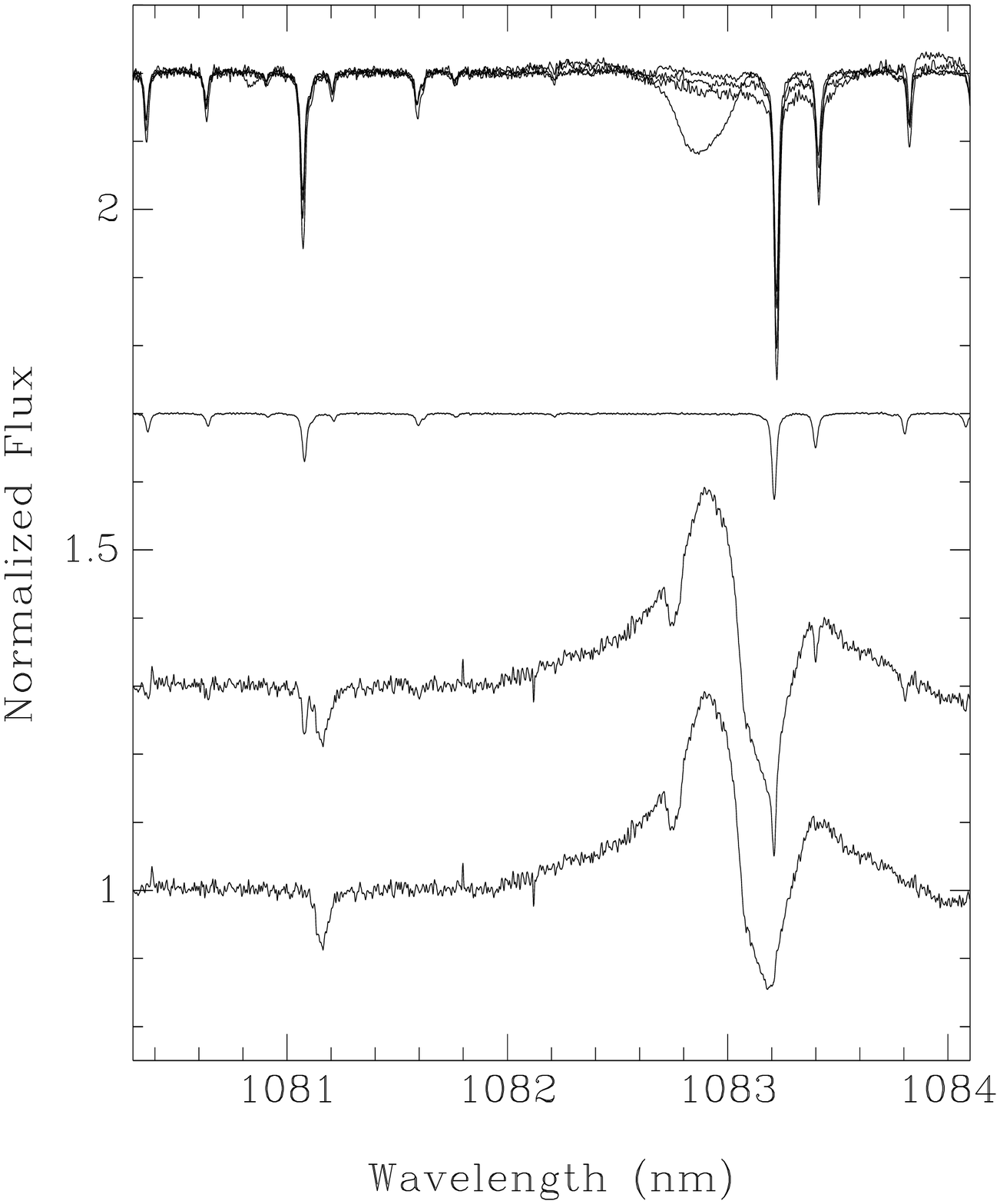}
\includegraphics[width=0.45\textwidth, height=7cm]{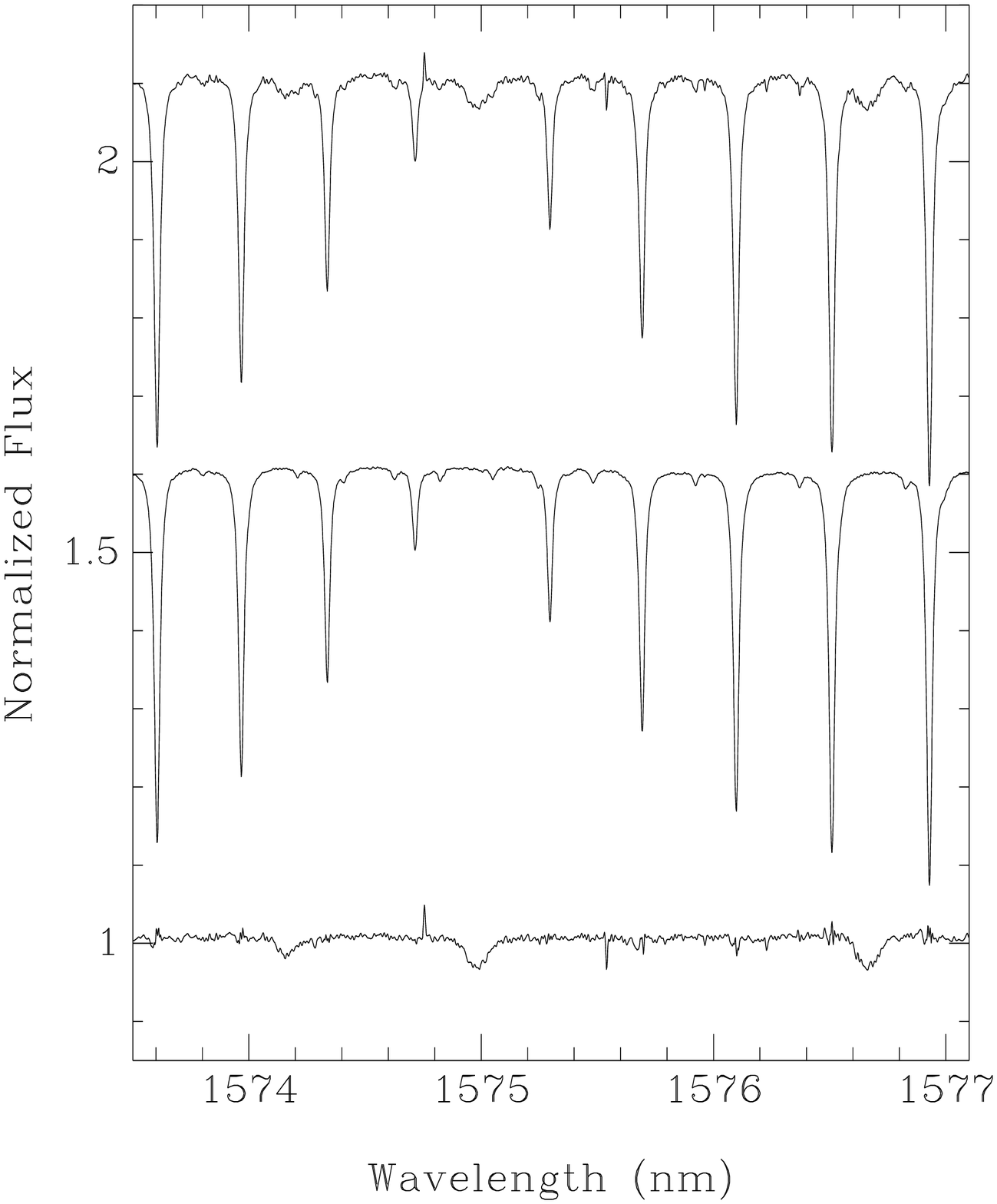}
\caption{Example for the removal of telluric lines in two spectral regions. 
Upper panel, from top to bottom:
spectra of four standard stars, scaled telluric spectrum, 
object (=HD\,101412) spectrum , and cleaned object spectrum.
Lower panel, from top to bottom: object (=HD\,101412) spectrum, telluric line master spectrum
(calculated by averaging seven standard star spectra), object spectrum
after removing the telluric lines.}
\label{fig:tell}
\end{figure}

Telluric features were removed in a few wavelength regions in the spectra of the Herbig Ae/Be star
HD\,101412 to search for magnetic splitting in a few \ion{Fe}{i} lines and to study the spectral
variability of line profiles belonging to different elements. After each science observation,
a hot rapidly rotating standard star has been observed at a comparable
zenith distance.
After moving the wavelength scale to the topocentric frame 
of rest, we averaged the normalized spectra of several standard stars 
to calculate a master spectrum of the telluric lines.
In order to match the telluric line intensity of each object spectrum,
this master telluric spectrum was scaled by increasing it to an 
appropriate power. Then, each object spectrum was divided by
its telluric spectrum to obtain a spectrum essentially free of 
telluric lines.
In this way, the corrected object spectrum $S_{\rm corr}$ was calculated from
the observed spectrum $S_{\rm obs}$ and the telluric line master spectrum $T$
as:
$$
S_{\rm corr} = S_{\rm obs} \cdot T^{-\alpha},
$$
where  $\alpha$ is the ratio of optical thickness of the terrestrial
atmosphere in the spectra $S_{\rm obs}$ and $T$, respectively. 
The coefficient $\alpha$ was chosen to obtain a flat spectrum in the regions
without stellar lines. Example for the removal of telluric lines in two spectral regions is 
presented in Fig.~\ref{fig:tell}.

%\section{Input line lists and synthetic spectra}
\section{The line identification}

The stellar lines were identified using the method of spectrum synthesis.
For each star, we computed an ATLAS9 model with parameters taken
from previous studies carried out by various authors: we adopted
$\teff=7700\,K$, $\logg=4.2$ for $\gamma$\,Equ 
(Ryabchikova et al.\ \cite{Ryabchikova1997}), $\teff=6800\,K$, $\logg=4.11$
for HD\,154708 (Nesvacil et al.\ \cite{Nesvacil2008}), $\teff=8300\,K$,
$\logg=3.8$  for HD\,101412 (Cowley et al.\ \cite{Cowley2010}), and 
$\teff=10,250\,K$, $\logg=3.6$ for 51\,Oph
 (Montesinos et al.\ \cite{Montesinos2009}), where $g$ is in cgs units. 
 
The models were used to compute synthetic spectra with the
\textsc{SYNTHE} code  (Kurucz \cite{Kurucz1993}). We adopted the atomic line lists
taken from Kurucz's web site (http://kurucz.harvard.edu/), but we substituted the
$\log\,gf$ values with those from the NIST database, whenever they 
were available. In addition, in some cases, we replaced the Kurucz $\log\,gf$ values for \ion{Si}{i} by those
from Mel\'endez \& Barbuy (\cite{Mel1999}).
% where available.
We added a few lines of \ion{Ce}{iii} 
with wavelengths and $\log\,gf$ values computed by
Bi\'emont (private communication), and a \ion{Dy}{ii} line at 
10835.94\,\AA\ taken from the VALD database (Kupka et al.\ \cite{Kupka1999};
Heiter et al.\ \cite{Heiter2008}). The
line broadening parameters are those computed by Kurucz; they are 
available for most of the identified lines, except for  the
\ion{Sr}{ii}, \ion{Ce}{iii}, and \ion{Dy}{ii} lines. where they were computed using classical 
approximations.

We started assuming solar abundances for all the elements.  
The synthetic spectra were broadened for an instrumental
resolving power of 95,000 and for
rotational velocities, which were determined from
the comparison of a few observed and computed profiles.
We found a $v\,\sin\,i$ of 16\,km\,s$^{-1}$ for $\gamma$\,Equ,
6\,km\,s$^{-1}$ for HD\,154708, 
9\,km\,s$^{-1}$ for HD\,101412, and
100\,km\,s$^{-1}$ for 51\,Oph.
These values are rather uncertain, owing to the 
magnetic fields that affect the line profile widths and that were neglected
in the computations.

For identification purposes, the abundances were modified
until agreement between the selected observed and computed profiles was achieved.
The abundances we obtained from the CRIRES spectra are only rough estimates
because, on one side, no Zeeman effect was considered in the line 
profile computations, and, on the other side, the spectra are affected 
by telluric lines, by 
several artifacts, by a wavelength calibration not fully correct.
Furthermore, the spectra  include entirely two hydrogen lines, 
at 1093.8086\,nm and 
1555.6467\,nm, and partially three other hydrogen lines: the extreme red
wing of 1543.8938\,nm, the red
wing of 1570.068\,nm,
and the blue wing of  1588.0558\,nm. 
Owing to the breadth of the hydrogen lines and the rather short 
($\sim$5\,nm) spectral regions covered by each of CRIRES detectors
it is very difficult to state were the continuum
level has to be placed almost everywhere in the observed spectra.  
However, in general, the element abundances derived from CRIRES
show a satisfactory agreement with the results of abundance determinations 
presented in 
previous studies using optical wavelengths.
Table~\ref{tab:abundances} lists the abundances adopted for computing the
synthetic spectra. For $\gamma$~Equ, HD\,154708, and HD\,101412 they are
 compared with the results from optical regions published by
Ryabchikova et al.\ (\cite{Ryabchikova1997}),
Nesvacil et al.\ (\cite{Nesvacil2008}),
and  Cowley et al.\ (\cite{Cowley2010}), respectively.

\begin{table*}
\caption{
The adopted abundances $\log(N_{\rm elem}/N_{\rm tot})$ and comparison with results
from the optical analyses by
Ryabchikova et al.\ (\cite{Ryabchikova1997}, RAW),
Nesvacil et al.\ (\cite{Nesvacil2008}, NHK), and
Cowley et al.\ (\cite{Cowley2010}, CHGS). A colon after the abundance value indicates
an uncertain abundance.
}
\begin{flushleft}
\label{tab:abundances}
\centering
\begin{tabular}{llllllllllllll}
\hline
\hline
\multicolumn{1}{c}{Element} &
\multicolumn{2}{c}{$\gamma$\,Equ} &
\multicolumn{2}{c}{HD\,154708} &
\multicolumn{2}{c}{HD\,101412} &
\multicolumn{1}{c}{51\,Oph} &
\multicolumn{1}{c}{} \\
&this work  & RAW&this work  & NHK  &this work  & CHGS & this work\\
\hline
\ion{He}{I} &       &        &        &  &        &       & $-$0.52  & \\
\ion{C}{i} & $-$4.0 & $-$3.66& $-$4.7:&  & $-$3.8 &$-$3.7          & \\
\ion{N}{i} &        &        &        &  & $-$3.5 & $-$3.49      & \\
\ion{O}{i} & $\le$ $-$3.7 &           &   &  & \\
\ion{Na}{i}& $\le$ $-$6.0 &        &$\le$ $-$6.7 & \\
\ion{Mg}{i}& $-$4.46 & $-$4.50& $-$5.8 & & $-$5.3 &$-$5.00 & $-$3.46\\
\ion{Al}{i}& $-$5.1 & $-$4.93&$\le$ $-$7.6\\
\ion{Si}{i}& $-$3.9 & $-$4.42& $-$5.6 & & $-$5.55 &$-$4.92 & $-$5.5\\
%           & $-$4.35\\
\ion{P}{i} & $\le$$-$7.8  &        &  \\
\ion{Ca}{i}& $-$5.1 & $-$5.40 &   $-$6.4 & $-$6.21\\
\ion{Ti}{i}&        &         & $\le$$-$7.7\\ 
\ion{Cr}{i}& $-$5.4 & $-$5.43 & $-$6.5 & $-$6.3\\
\ion{Fe}{i}& $-$4.35 & $-$4.28&$-$6.1 & $-$5.73 &$-$5.1 &$-$5.07 & $-$4.0 ?\\
\ion{Ni}{i}& $-$5.8 &          &$\le$$-$7.5 &$-$6.71\\
\ion{Sr}{ii}& $-$7.1 & $-$7.43& $-$9.07&\\
\ion{Ce}{ii}&    &$-$9.2 &  & $-$9.09\\       
\ion{Ce}{iii}$^{a}$ & $-$5.1 &       &$-$6.8\\
                  & $-$7.0 &      &  $-$7.2 \\
\ion{Dy}{ii}      &        & &$-$8.4 &\\
\hline
\end{tabular}
\end{flushleft}
%$^{a}$ Some \ion{Si}{i} lines in $\gamma$Equ give $-$3.9\,dex, other 
%$\ion{Si}{i} lines give $-$4.35\,dex\\
$^{a}$ the line at 1584.76\,nm gives higher abundances than other two lines 
at 1570.9638\,nm and 1571.5837\,nm. The relative abundance is 
 listed  in the subsequent row.\\
\end{table*}

The lists of identified lines including observed and laboratory wavelengths, 
central line depths, oscillator strengths, excitation potentials for the 
lower and upper levels of
the respective transition, and 
sources for the atomic data are presented in Tables~\ref{tab:a1}--\ref{tab:a3}.
We note that several spectral lines  
remain unidentified due to unavailability of atomic data. 
The identification list is most 
complete for the Ap star $\gamma$\,Equ and the Herbig Ae star HD\,101412.
The results of the line identification for
each target are discussed below in Sect.~\ref{sect:individual}.

The observed and synthetic spectra are presented on Castelli's web 
page\footnote{http://wwwuser.oat.ts.astro.it/castelli/stars/}
together with the line-by-line identification for each studied star.

\section{Individual targets}
\label{sect:individual}

\subsection{$\gamma$\,Equ}

The classical Ap star $\gamma$\,Equ is one of the best studied Ap stars for which the presence
of a magnetic field was discovered several decades ago (Babcock \cite{Babcock1958}).
Due to its brightness, slow rotation, and the presence of a rather strong surface magnetic field of the order of 4\,kG
this star is frequently used
as a standard star in spectropolarimetric observations. Broad-band oscillations with a period
of 12.4\,min were discovered by Kurtz (\cite{Kurtz1983}) and radial velocity 
oscillations were first studied by Libbrecht (\cite{Libb1988}). $\gamma$\,Equ belongs to the group of 
rapidly oscillating Ap (roAp) stars, which are main-sequence stars that pulsate in high radial 
overtone p modes with periods in 
the range of 5.7--21\,min. RoAp stars show broad-band photometric amplitudes less than 0.01\,mag, 
whereas rapid radial velocity variations in rare earth element lines can reach 8\,km\,s$^{-1}$
(e.g. Kurtz \cite{Kurtz1990}: Freyhammer et al.\ \cite{Freyhammer2009}). 

Among the studied targets, the list of identified lines is the longest and most complete for $\gamma$\,Equ.
The definitely identified lines include \ion{H}{i}, \ion{C}{i}, \ion{Si}{i}, \ion{Ca}{i}, 
\ion{Mg}{i}, \ion{Mg}{ii}, \ion{Cr}{i}, \ion{Fe}{i}, \ion{Ni}{i}, \ion{Sr}{ii},
and  two lines belonging to the rare-earth element group, \ion{Ce}{iii} 1584.76\,nm and 
\ion{Dy}{ii} 1083.59\,nm.
An additional \ion{Ce}{iii} line at 1571.5837\,nm could be present in the wing of
a telluric line. 
The presence of titanium is doubtful, because we possibly identified two 
\ion{Ti}{ii} lines at 1073.374\,nm and 1077.1294\,nm, but no \ion{Ti}{i}
lines.
All identified lines in the spectra of this target are presented in Table~\ref{tab:a1}.

For the \ion{H}{i} lines at 1093.8086\,nm and 1555.6467\,nm, the core and both wings are
contained in the observed range. For the line at 1570.068\,nm we see the red wing, for 
that at 1588.0558\,nm we see the blue wing. 

The only element for which the abundance can be rather well established is \ion{Mg}{i}.
The solar abundance was determined from the line at 1081.11\,nm. It well reproduces
other identified \ion{Mg}{i} lines, except for those lying on the red wing of
\ion{H}{i} 1093.8\,nm. The only well observed \ion{Mg}{ii} line at 1095.1778\,nm 
is poorly reproduced. We note that the \ion{Mg}{i} line at 1081.11\, nm is subject to 
full Paschen-Back effect. A more complete discussion of this line is presented in the next 
subsection.

Most of the \ion{Fe}{i} lines in the 1550.0-1585.0\,nm interval are
split in three components by the magnetic field, so that they were not used
for abundance purposes as they cannot be properly dealt with using the SYNTHE code.
The iron abundance of $-$4.35\,dex was  derived from the blend \ion{Fe}{i}, \ion{Fe}{ii}
at 1087.16\,nm and from the unsplit (or weakly split) \ion{Fe}{i} lines
at 1576.942\,nm and 1577.407\,nm. 

When we analyzed the \ion{Si}{i} lines, we derived two different \ion{Si}{i} abundances.
A first group of lines was rather well reproduced by  $-$3.9\,dex, as derived
from \ion{Si}{i} at 1074.9378\,nm, a second group of lines was well reproduced
by $-$4.35\,dex, as derived from the line at 1088.533\,nm. The source for  
the $\log\,gf$ values was NIST4 for
most lines of the first group and Kurucz (\cite{Kurucz2007}) for most lines of the second group.
The disagreement disappeared after the Kurucz (\cite{Kurucz2007}) $\log\,gf$ values were replaced by
Mel\'endez \& Barbuy (\cite{Mel1999}) $\log\,gf$ data, as suggested by the referee. In this way,
the whole \ion{Si}{i} observed spectrum is rather well synthetized by the sole abundance of $-$3.9\,dex.
The line \ion{Si}{i} at 1555.778\,nm is split into two components due to the presence 
of the magnetic field.
%
%The identified \ion{Si}{i} lines give rise to two different \ion{Si}{i} abundances.
%A first group of lines is rather well reproduced by  $-$3.9\,dex as derived
%from \ion{Si}{i} at 1074.9378\,nm, a second group of lines is well reproduced
%by $-$4.35 dex, as derived from the line at 1088.533\,nm. A possible explanation could be
%the use of different sources for the $\log\,gf$ values. The source is NIST4 for
%most lines of the first group and Kurucz (\cite{Kurucz2007}) 
%for most lines of the second group.
%The line \ion{Si}{i} at 1555.778\,nm is split into two components due to the presence 
%of the magnetic field.

All other elements are present with very few lines, usually weak, so that
their abundance was derived either from the only observed line (\ion{Al}{i}
at 1089.172\,nm, \ion{Ni}{i} at 1555.537\,nm, \ion{Sr}{ii} at 1091.4887\,nm)
or from that of the lines which we estimated as the best suited to give the abundance
(\ion{C}{i} at 1072.955\,nm, \ion{Ca}{i} at 1083.897\,nm, \ion{Cr}{i} at 1090.586\,nm, \ion{Ce}{iii} 
at 1584.755\,nm).
The abundances in these cases are very uncertain, as the Zeeman effect was not considered
in the line profile computations. In fact,
when other lines are present, they would have required somewhat different abundances.
A second \ion{Ce}{iii} line at 1571.5837 nm could be present in the wing 
of a telluric line. But it is predicted much stronger than observed for the 
abundance of -5.1\,dex given by the line at 1584.76\,nm. This is also the case 
for the \ion{Ce}{iii} line at 1570.9638\,nm, which is
even not observed. For these two lines, a lowering of the \ion{Ce}{iii} abundance 
up -7.0\,dex is 
needed to obtain agreement between the observed and computed spectra.
An exception is \ion{Cr}{i}, for which the two observed lines are both well
reproduced by the same abundance.

The abundances for \ion{O}{i}, \ion{Na}{i}, and \ion{P}{i} are upper limits, because there are
no lines of these elements observed in the spectrum, while they were predicted
for solar abundances (\ion{O}{i} at 1074.55\,nm, \ion{Na}{i} at 1083.487\,nm, 
\ion{P}{i} at 1076.9511\,nm).
 For titanium, whose presence is doubtful, we adopted solar abundance.
\begin{figure*}
\centering
\includegraphics[width=0.40\textwidth]{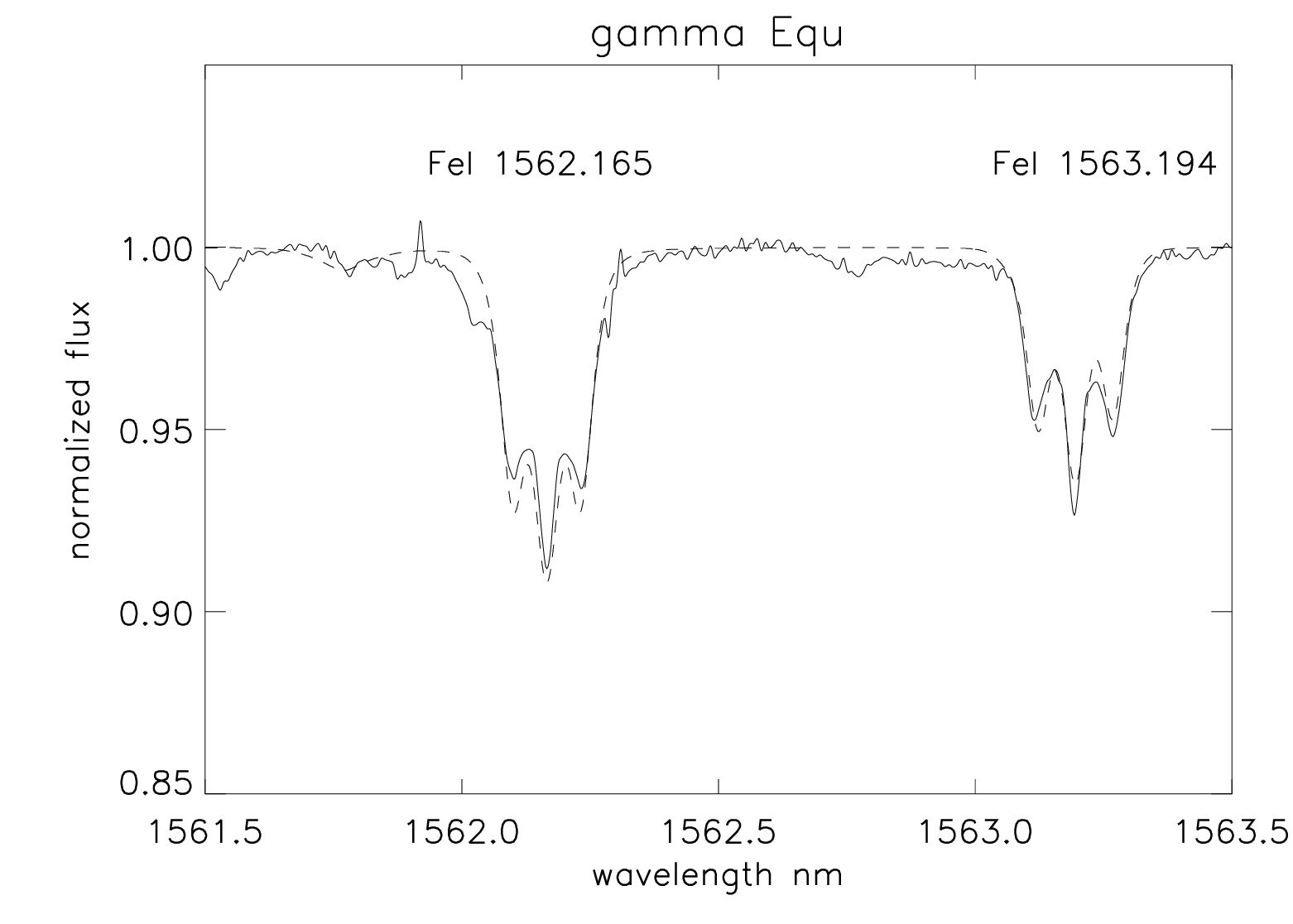}
\includegraphics[width=0.40\textwidth]{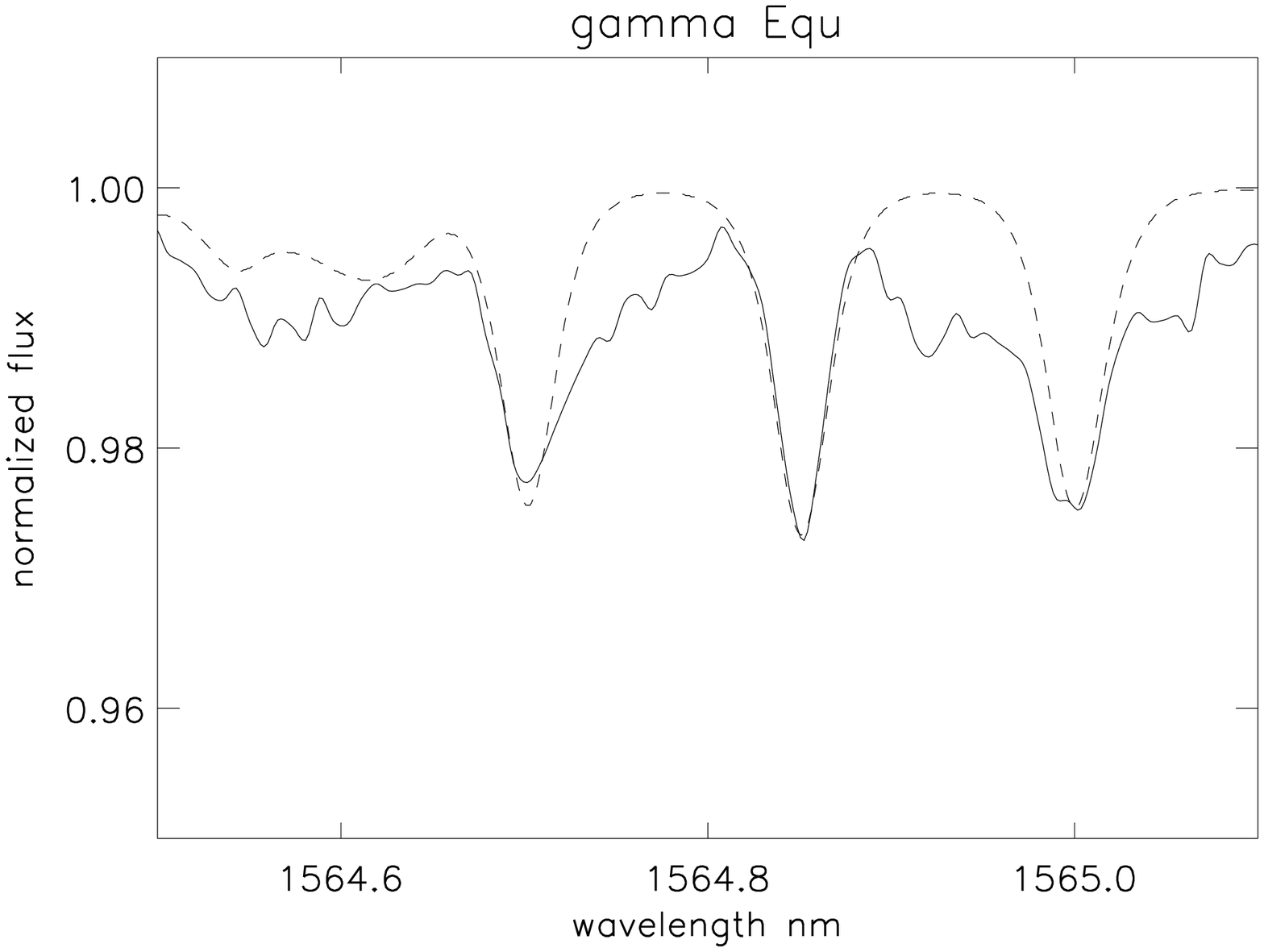}
\caption{
{\sl Left:}  Synthetic line profiles (dashed line) calculated for \ion{Fe}{i} 15621.7\,nm and 15631.9\,nm lines 
using the \textsc{SYNTHMAG} code.
{\sl Right:} Synthetic line profile (dashed line) for the \ion{Fe}{i} 15648.5\,nm line.
}
\label{fig:fe_15648}
\end{figure*}

We note that line identification in Ap stars with resolved Zeeman
split lines can be strengthened by comparing the observed and expected magnetic splitting 
patterns. In stars with strong fields, both the central line position and the whole line profile 
shape, as determined by the number and relative strengths of the $\pi $ and $\sigma $ components, 
can serve as a consistency check in the cases 
where line identification is doubtful. However, such a procedure implies 
the availability of Land\'e factors for the majority of the observed spectral lines in the 
near-IR wavelength region, 
which is not the case for the currently available atomic data bases.

Synthetic line profiles were also calculated for a few magnetically sensitive iron lines 
with known Land\'e factors and showing magnetically
split lines. For the synthesis we used the
software \textsc{SYNTHMAG} developed by Piskunov (\cite{Piskunov99}).
The results of our synthesis using this code assuming a surface magnetic field 
of 4.0\,kG, $v\,\sin\,i=0$\,km\,s,
and an iron abundance $-$4.4\,dex are presented in Fig.~\ref{fig:fe_15648}.

\subsection{HD\,154708}

This target possesses one of the strongest magnetic fields detected among the Ap stars 
(Hubrig et al.\ \cite{Hubrig2005,Hubrig2006}). 
The magnetic field modulus of HD\,154708 is very large, of the order of 25\,kG.
We note that stars with magnetic field strengths that exceed 20\,kG are very rare and 
only very few such strongly magnetic stars were detected so far.  
HD\,154708 has a luminosity and temperature that place it at the cool end of the range of
known roAp stars.  
Kurtz et al.\ (\cite{Kurtz2006}) obtained for this star time series of high-resolution 
spectra and showed that it indeed belongs to the group of roAp stars.
The pulsation takes place with a period of 8\,min and the measured radial velocity amplitudes for the rare 
earth ions \ion{Nd}{iii}, \ion{Pr}{ii}, and \ion{Pr}{iii} are of the order of 60\,m\,s$^{-1}$.

The magnetic field of HD\,154708 is so strong that in our previous studies we were not able to interpret the 
spectrum in certain optical regions. Many features corresponding to the known transitions 
were distorted beyond recognition by magnetic splitting.

The optical spectrum of HD\,154708 displays the typical abundance pattern of roAp stars 
with overabundance of rare earth elements along with their clear ionization imbalance, i.e.
the third spectra of the rare earths generally give higher abundances
by about 1\,dex than the second spectra. . 
Using Geneva photometry, Nesvacil et al.\ (\cite{Nesvacil2008}) estimated $\teff=6800\,K$
and $\logg=4.1$.
Their abundance analysis revealed that light elements, as well as Ti, Fe, and Ni are underabundant. 

\begin{figure}
\centering
\includegraphics[angle=90,width=0.40\textwidth]{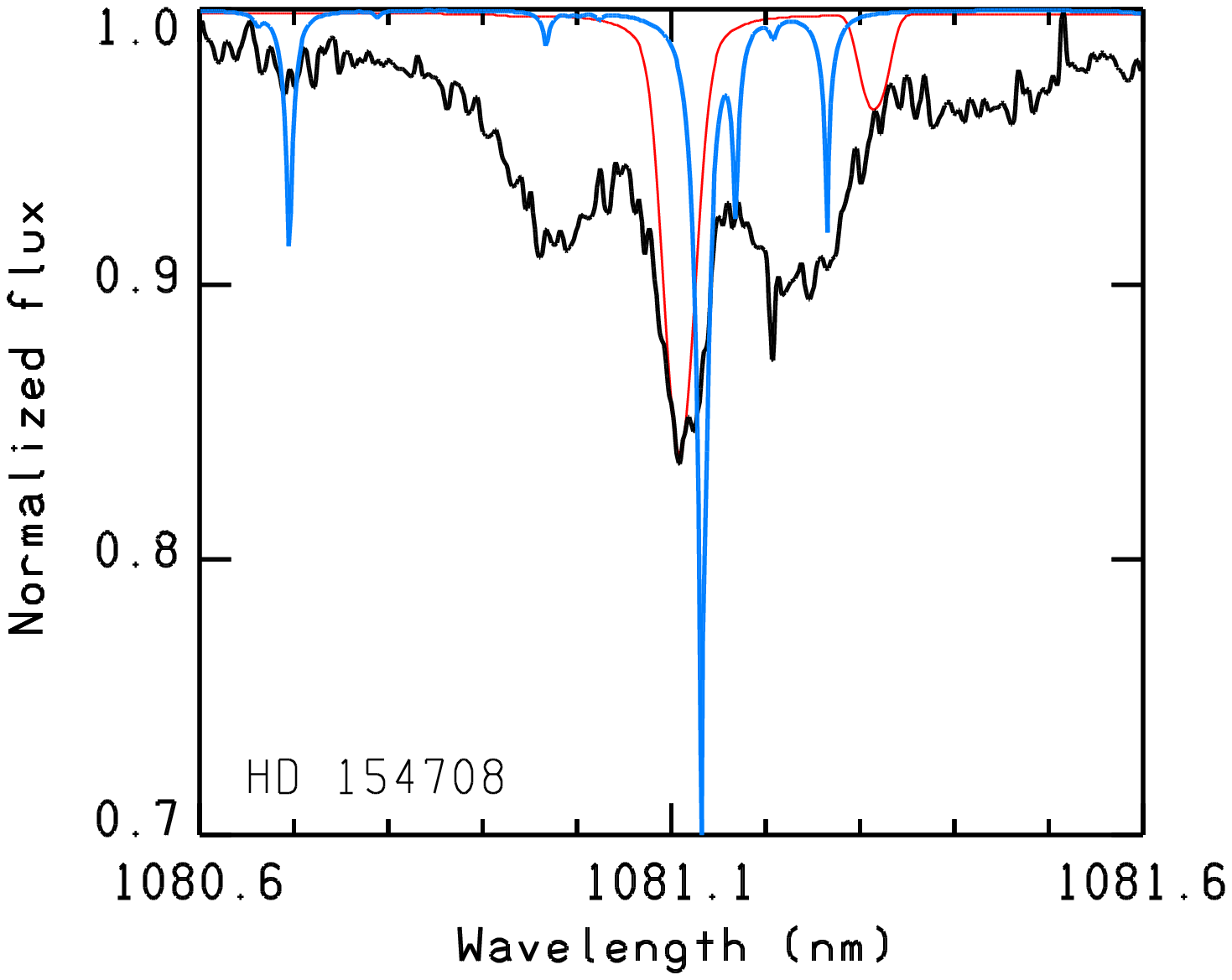}
\includegraphics[angle=90,width=0.40\textwidth]{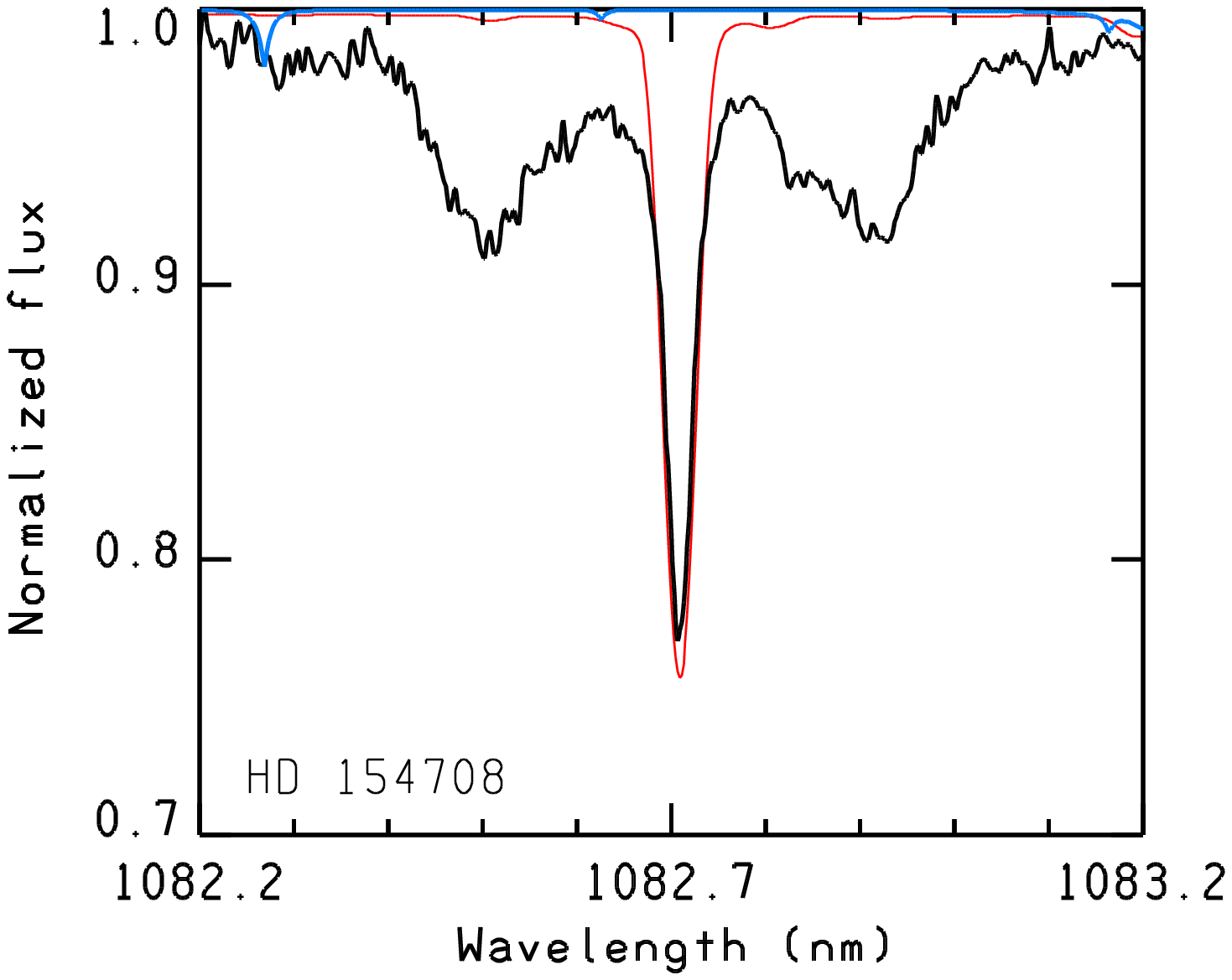}
\includegraphics[angle=90,width=0.40\textwidth]{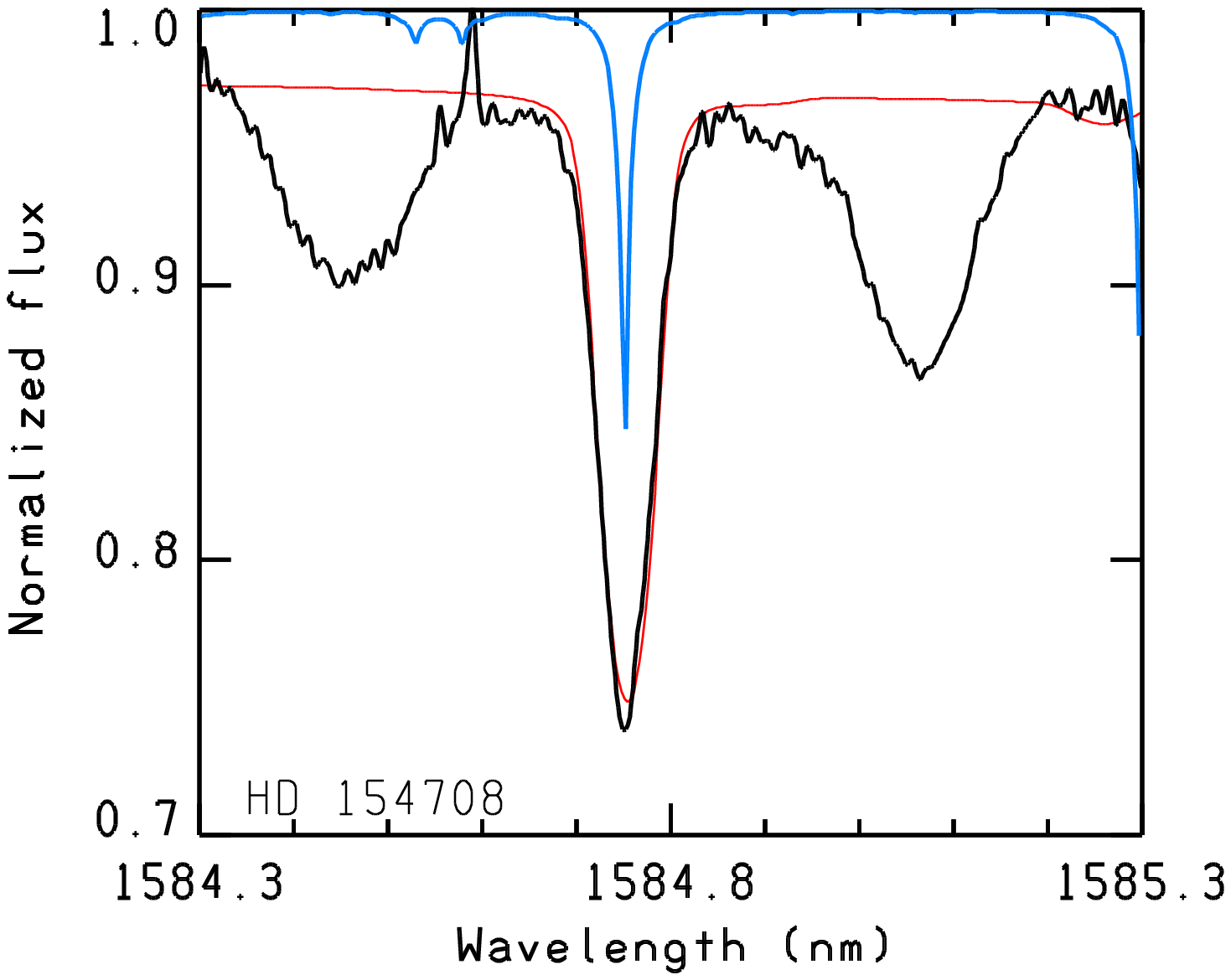}
\caption{Magnetically split lines belonging to \ion{Mg}{i} 1081.11\,nm, \ion{Si}{i} 1082.71\,nm,
and \ion{Ce}{iii} 1584.76\,nm in the CRIRES spectrum of HD\,154708. The blue lines in the online
version of the article indicate the contribution of the telluric absorptions. Only the central components have been fitted
in the line identification process.}
\label{fig:1083}
\end{figure}

Our line identification study indicates the presence of 
\ion{H}{i}, \ion{C}{i}, \ion{Mg}{i}, \ion{Mg}{ii}, \ion{Si}{i}, and \ion{Fe}{i} lines.
Also one \ion{Sr}{ii} line, and three
lines belonging to the rare-earth element group, \ion{Ce}{iii} 1571.58\,nm and  1584.76\,nm,
and \ion{Dy}{ii} 1083.594\,nm are identified. 
Nearly 30\% of the spectral lines remain unidentified both due to unavailability
of atomic data and to the complex structure of the profiles, which sometimes
are the blend of the central component of a line with the split component 
of a nearby line. 

Abundances are difficult to determine owing to the structure 
of the lines. We derived abundances by reproducing the central undisplaced component
of the observed profiles. We used  1081.11\,nm for \ion{Mg}{i}, 1074.938\,nm
for \ion{Si}{i}, and 1484.755\,nm for \ion{Ce}{iii}.
The \ion{Fe}{i} lines appear in the spectra unsplit.  
%but it is possible that we observe only the central component(s) due to the presence of  Paschen-Back effect.
For most of them the Paschen-Back effect should be at most moderate, if detectable at all. 
Furthermore, in their vast majority, they have Zeeman patterns that are very close to pure triplets.
As a matetr of fact, it is puzzling why we do not observe the magnetic splitting of these lines. 
The iron abundance was derived from the
line at 1562.16\,nm. 
The abundances of the other elements were obtained from the respectively only
line observed (\ion{C}{i} at 1572.376\,nm, \ion{Ca}{i} at 1083.897\,nm,
\ion{Cr}{i} at 1101.5679\,nm, \ion{Sr}{ii} at 1091.4887\,nm).

The abundances for \ion{Na}{i}, \ion{Al}{i}, \ion{Ti}{i}, \ion{Ni}{i} are
upper limits because there are no lines observed for them, although they were
predicted for solar abundances (\ion{Na}{i} at 1083.487\,nm, 
\ion{Al}{i} at 1089.172\,nm, \ion{Ti}{i} at 1554.3758\,nm, and
\ion{Ni}{i} at 1555.537\,nm). Nothing can be said about \ion{O}{i}  
because a strong artifact overlaps the \ion{O}{i} line at 1074.55\,nm.

In Fig.~\ref{fig:1083} we present lines of the magnetically split lines belonging to \ion{Mg}{i} 1081.11\,nm,
\ion{Si}{i} 1082.71\,nm, and \ion{Ce}{iii} 1584.76\,nm showing a rather similar 
split Zeeman structure. 
%The \ion{Mg}{i} 10811.1\,nm line is actually a blend of five \ion{Mg}{i} lines,
%where the 1081.1053\,nm line is the strongest. 
The structure of the displayed \ion{Ce}{iii} 1584.76\,nm line, which splits into
three components, is very similar to that of the
\ion{Nd}{iii} 6245.1\,\AA{} line previously observed in the optical UVES spectrum by Nesvacil et al.\
(\cite{Nesvacil2008}).
Numerous examples of magnetically split lines are presented under the link
http://wwwuser.oat.ts.astro.it/castelli/stars/.

\begin{table*}
\caption{
Examples of a few transitions of spectral lines showing magnetic splitting.
}
\label{tab:split}
\centering 
\begin{tabular}{llllllllllllll}
\hline
\hline
\multicolumn{1}{c}{Element{\Rv}} &
\multicolumn{1}{c}{$\lambda$(lab)} &
\multicolumn{1}{c}{$\log\,gf$} &
\multicolumn{1}{c}{$\chi_{\rm low}$} &
\multicolumn{1}{c}{Level{\Rr}} &
\multicolumn{1}{c}{$\chi_{\rm up}$} &
\multicolumn{1}{c}{Level{\Rr}} \\
\multicolumn{1}{c}{  } &
\multicolumn{1}{c}{[nm]{\Rb}} &
\multicolumn{1}{c}{ } &
\multicolumn{1}{c}{[cm$^{-1}$]} &
\multicolumn{1}{c}{{\Rh}Desig.{\Rr}} &
\multicolumn{1}{c}{[cm$^{-1}$]} &
\multicolumn{1}{c}{{\Rh}Desig.{\Rr}}
\\
\hline
\ion{Mg{\Rv}}{i} & 1081.1053 &{\Rh} 0.024& 47957.045 & 3s 3d  $\rm ^3D_3$ & 57204.305 & 3s5f $\rm ^3F^o_4$\\
\ion{Si}{i} & 1082.7089 &{\Rh} 0.239& 39955.053 & 3s$^2$ 3p4s $\rm ^3P^o_2$ & 49188.617 & 3s$^2$3p4p $\rm ^3P_2$ \\
\ion{Si}{i} & 1086.8790 &{\Rh} 0.206& 49933.775 & 3s$^2$ 3p3d $\rm ^3F^o_3$ & 59131.912 & 3s$^2$3p4f $^2[9/2]_4$\\
\ion{Sr}{ii}& 1091.4887 &$-$0.638 &14555.900 & 4d $\rm ^2D_{3/2}$ & 23715.19 & 5p $\rm ^2P^o_{1/2}$ \\
\hline
\end{tabular} 
\end{table*} 

In Table~\ref{tab:split} we present the transitions for a few selected lines showing magnetic splitting.
In the Columns 1 to 7 we list laboratory wavelengths, 
oscillator strengths, excitation potentials for the 
lower and upper levels, and designations of lower and upper terms.

\begin{figure}
\centering
\includegraphics[width=0.50\textwidth]{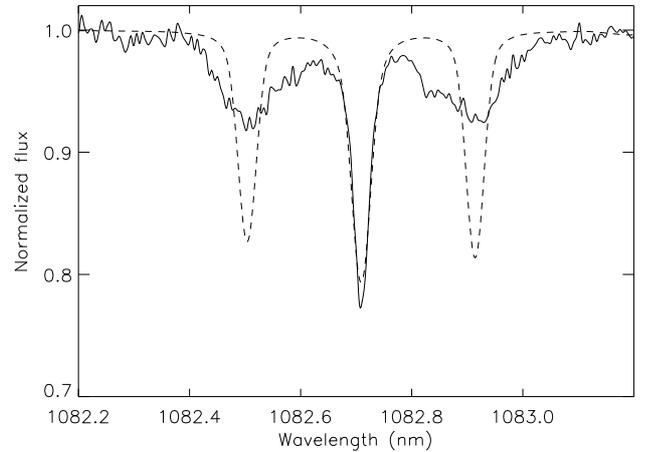}
\caption{
Synthetic line profile (dashed line) calculated for \ion{Si}{i} 1082.71\,nm 
in the spectrum of HD\,154708 using the \textsc{SYNTHMAG} code.
}
\label{fig:si_syn}
\end{figure}

Our attempt to model the magnetically split lines is presented in Fig.~\ref{fig:si_syn},
where we tried to calculate the synthetic profile of the line \ion{Si}{i} 1082.71\,nm using 
the \textsc{SYNTHMAG} code.
The interesting feature of the split line profiles in the spectra of HD\,154708 is that the 
$\sigma$ components are broad, in particular considerably broader than the $\pi$ components. 
This indicates that the spread of the field strengths over the visible stellar hemisphere is 
rather large, probably significantly larger than it would be for a centred dipole. It would be 
important in the future to prove whether the width of the $\sigma$ components varies with
rotation phase.
The splitting of all levels involved in most Si transitions should not significantly 
depart from the linear Zeeman effect. Furthermore, the Zeeman patterns of most transitions are pure 
triplets. For the split line \ion{Si}{i} 1084.39\,nm, assuming LS coupling, the Land\'e factor is 1.0; 
for the transitions \ion{Si}{i} 1074.94\,nm, 1082.71\,nm, and 1097.93\,nm it is 1.5. From the plot of the 
line at 1082.71\,nm we estimate that the two $\sigma$ components are separated by 0.41\,nm. 
For a triplet with $g=1.5$, this corresponds to a mean field modulus of 25.0\,kG. This result is in 
agreement with the field modulus determinations from the visible. 

On the other hand, with a field of 25\,kG or more, the upper level of the \ion{Si}{i} 
transition at 1086.88\,nm
is subject to partial Paschen-Back effect, while the linear Zeeman effect remains an excellent 
approximation for the lower level. Thus one would expect to see some distortion of the pattern, 
similar to what happens for the \ion{Fe}{ii} lines at 6147.7--6149.2\,\AA{} in the visual 
spectrum of this star. 
%%Interestingly, the occurrence of partial Paschen-Back effect should 
%%also make the transition between the same lower level and the other level of the upper 
%%term, 59128.40, an allowed transition. (Nominally, it is a forbidden transition, J=3 -> J=5.)

In the absence of a magnetic field, the line at 1081.11\,nm results from the superposition of five 
transitions of \ion{Mg}{i}, of which the one at 1081.1053\,nm is the strongest. The lower levels of 
these transitions all belong to the term $3s3d~^3$D, and all the upper levels belong to the 
term $3s5f~^3$F$^{\circ}$. The individual transitions correspond to different combinations of the lower 
and upper $J$ quantum numbers. Within both the lower and the upper term, the levels of different 
$J$ are separated from each other by less than 0.1~cm$^{-1}$. Thus in the strong magnetic 
field of HD\,154708, both terms are subject to full Paschen-Back effect, and $J$ ceases to be 
a "good" quantum number. As a result, the observed line at 1081.11\,nm really originates from a single 
transition between the two terms of interest, which is split by the magnetic field with a pure 
triplet pattern.

%The identified \ion{Mg}{i} transitions are subject to full Paschen-Back effect. It no longer 
%makes sense to distinguish between the levels of different J within each of the involved terms. 
%LS coupling no longer applies. 
%%The three lines that you list are actually single transitions, rather than the superposition of several different transitions. I believe that the Landé factor in this case should be taken equal to 1.0, which doe
%%s not seem inconsistent with the observed splitting. (I should double check on this, though.)

The linear Zeeman effect applies for both levels in the transition \ion{Sr}{ii} 1091.49\,nm. 
The Zeeman pattern closely resembles a triplet, as separations are small within the pairs of the
individual $\pi$ components and of the individual $\sigma$ components. 

As for the rare earth \ion{Ce}{iii} and \ion{Dy}{ii} lines, they most frequently display Zeeman patterns
%Without level information, one cannot say much about the structure of the magnetically split line
%\ion{Ce}{iii} at ~1584.76\,nm. However, most frequent for rare earths is to have Zeeman patterns 
that do not depart much from triplets, with Land\'e factors fairly close to 1. The observed 
splitting of the \ion{Ce}{iii} and \ion{Dy}{ii} lines is not inconsistent with this. 
Admittedly, the red $\sigma$ 
component of the line \ion{Ce}{iii} at 1584.76\,nm appears deeper and narrower than the blue one. 
We are not sure at present how to explain it since there is absolutely no reason why a departure 
from linear Zeeman effect should be observable in this line.
%which is suggestive of partial Paschen-Back effect. 
We note again that the observed $\sigma$ components are 
considerably broader than the observed $\pi$ components, consistent with the previous conclusion 
about the broad range of values of the field strengths across the stellar disk. 
A more detailed study of Zeeman splittings observed in this strongly magnetic star is in process
and will be presented in a separate paper.
The list of all identified lines is presented in Table~\ref{tab:a2}.

\subsection{HD\,101412}

Among the sample of Herbig Ae/Be stars with detected magnetic fields, HD\,101412 showed the 
largest longitudinal magnetic 
field, $\left<B_{\rm z}\right>$\,=\,$-$454$\pm$42\,G, measured on low-resolution polarimetric spectra
obtained with FORS\,1 (FOcal Reducer 
low-dispersion Spectrograph) mounted on the 8-m Kueyen (UT2) telescope of the VLT (Hubrig et al.\
\cite{Hubrig2009b}).
The subsequent study of twelve UVES and HARPS spectra of HD\,101412 revealed the presence of 
resolved magnetically 
split lines indicating a variable magnetic field modulus changing  from 2.5 to 3.5\,kG (Hubrig et al.\
\cite{Hubrig2010}).
The search of the rotation period using magnetic and photometric data 
resulted in  $P=42.076 \pm 0.017$\,d (Hubrig et al.\ \cite{Hubrig2011}). 
The presence of a rather strong magnetic field on the surface of HD\,101412 makes it one of the best
candidates for studies of the impact of the magnetic field on the physical processes occurring 
during stellar formation.

Our previous study of the abundances of HD\,101412 using UVES and HARPS spectra
indicated that they may reflect a mild $\lambda$\,Boo, or Vega-like abundance mechanism, 
where the refractory elements are depleted while the most volatile elements are 
nearly normal (Cowley et al.\ \cite{Cowley2010}). The inspection of the spectroscopic material 
in the visual wavelength region
revealed the presence of the elements He, C, N, O, Na, Mg, Al, Si, S, Ca, Sc, Ti, V, Cr, Mn, Fe, 
Co, Ni, Zn, Sr, Y, Zr, and Ba. 
Almost all spectral lines showed variations in line intensity and line profile, with the most
pronounced variability detected for lines of the elements
He, Si, Mg, Ca, Ti, Cr, Fe Sr, Y, Zr, and Ba. Since also the magnetically insensitive lines 
showed clear profile variations,
we concluded that the detected spectral variability is a combination of both Zeeman splitting 
and abundance spots (Hubrig et al.\ \cite{Hubrig2010}).

\begin{figure}
\centering
\includegraphics[width=0.45\textwidth]{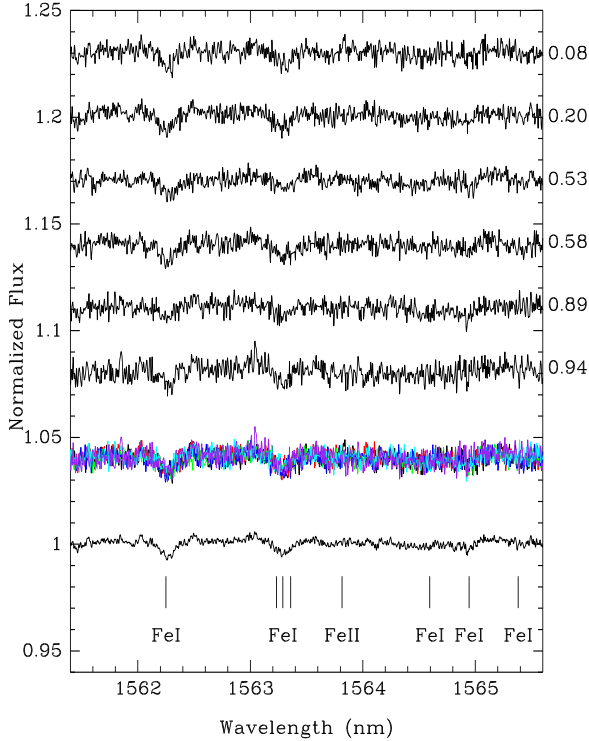}
\caption{Magnetically sensitive \ion{Fe}{i} lines at 15621.7\,nm and 15631.9\,nm in the 
CRIRES spectra of HD\,101412 after the 
removal of telluric lines and correction for heliocentric velocity.
The spectrum at the bottom displays the average of the six individual spectra obtained at different
rotation phases. In the second spectrum from the bottom all six spectra are overplotted. 
}
\label{fig:15640}
\end{figure}
 
\begin{figure}
\centering
\includegraphics[width=0.40\textwidth]{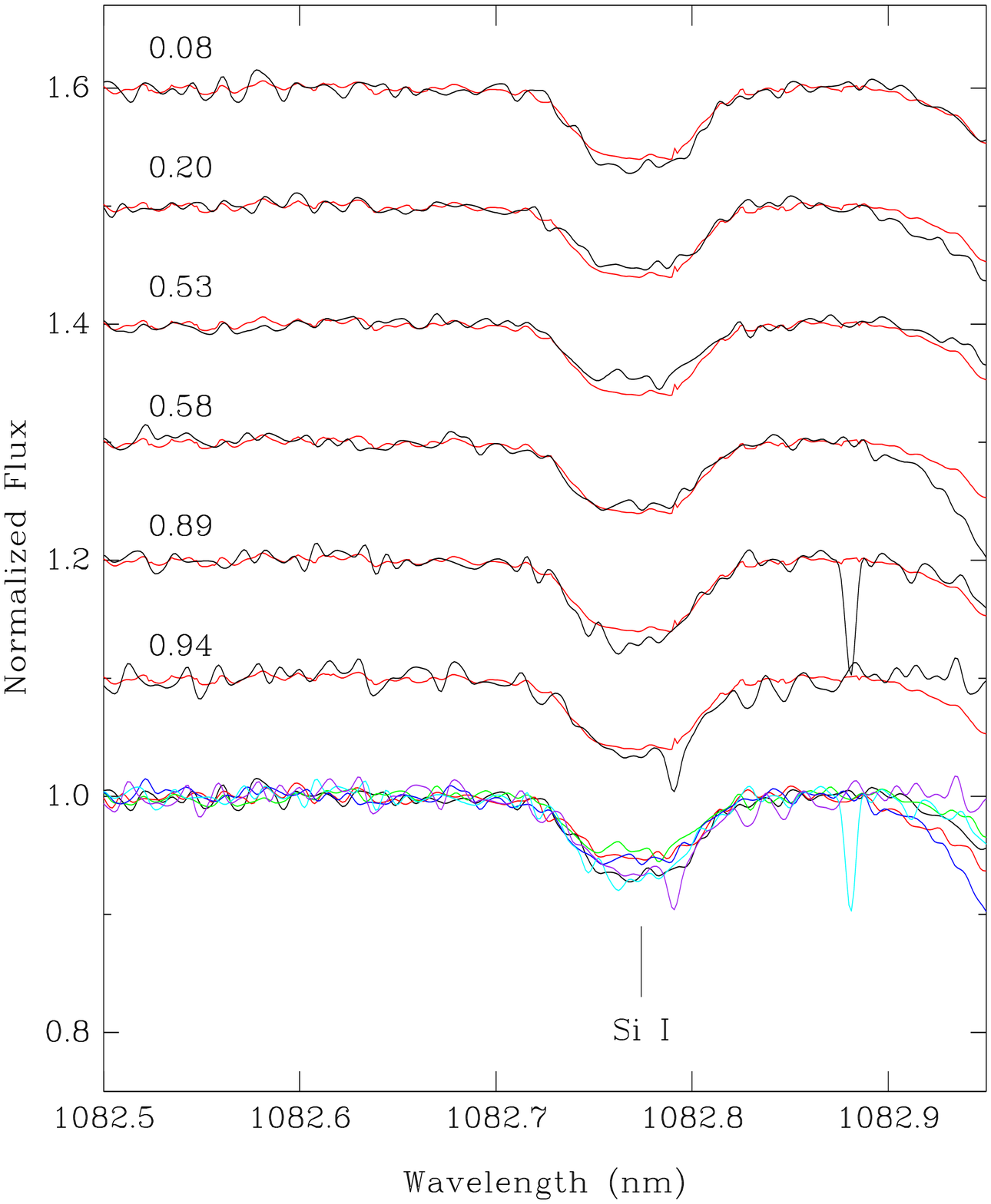}
\caption{
Line profile variations detected in the \ion{Si}{i} 1082.7\,nm line over the rotation period in the 
spectra of HD\,101412. The smooth overplotted lines (in red colour in the online version)
present the average spectrum.
The spectrum at the bottom of the figure presents all six spectra overplotted.}
\label{fig:Sinorm}
\end{figure}

\begin{figure}
\centering
\includegraphics[width=0.40\textwidth]{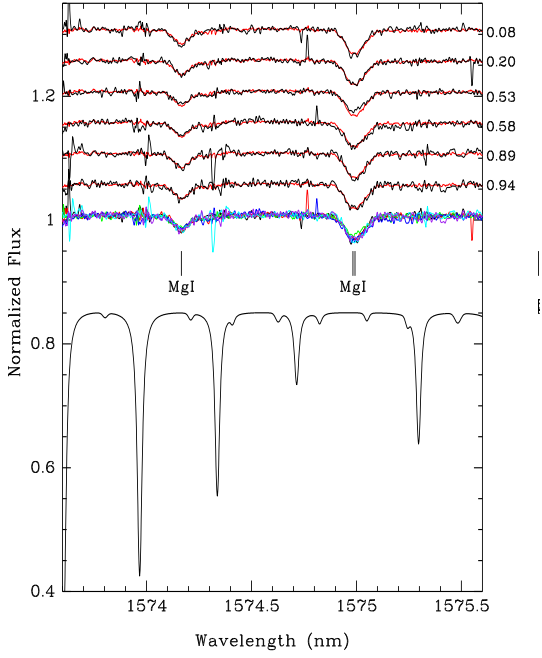}
\caption{
Line profile variations detected in the \ion{Mg}{i} 1574.1\,nm and 1574.9\,nm lines over the 
rotation period in the spectra of HD\,101412. The smooth overplotted lines (in red colour in the online version)
present the average spectrum.
The telluric spectrum presented at the bottom of the figure was removed from the observed spectrum.
The second spectrum from the bottom presents all six spectra overplotted.  
}
\label{fig:1574Mgclean}
\end{figure}

\begin{figure}
\centering
\includegraphics[width=0.40\textwidth]{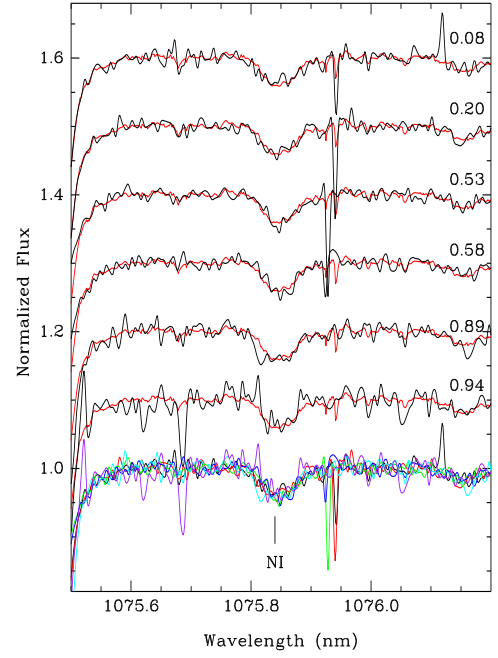}
\caption{
Behaviour of the line profile of \ion{N}{i} 1075.8\,nm over the rotation period
in the spectra of HD\,101412. The smooth overplotted lines (in red colour in the online version)
present the average spectrum.
The spectrum at the bottom of the figure presents all six spectra overplotted.
}
\label{fig:1076NI}
\end{figure}

\begin{figure}
\centering
\includegraphics[width=0.40\textwidth]{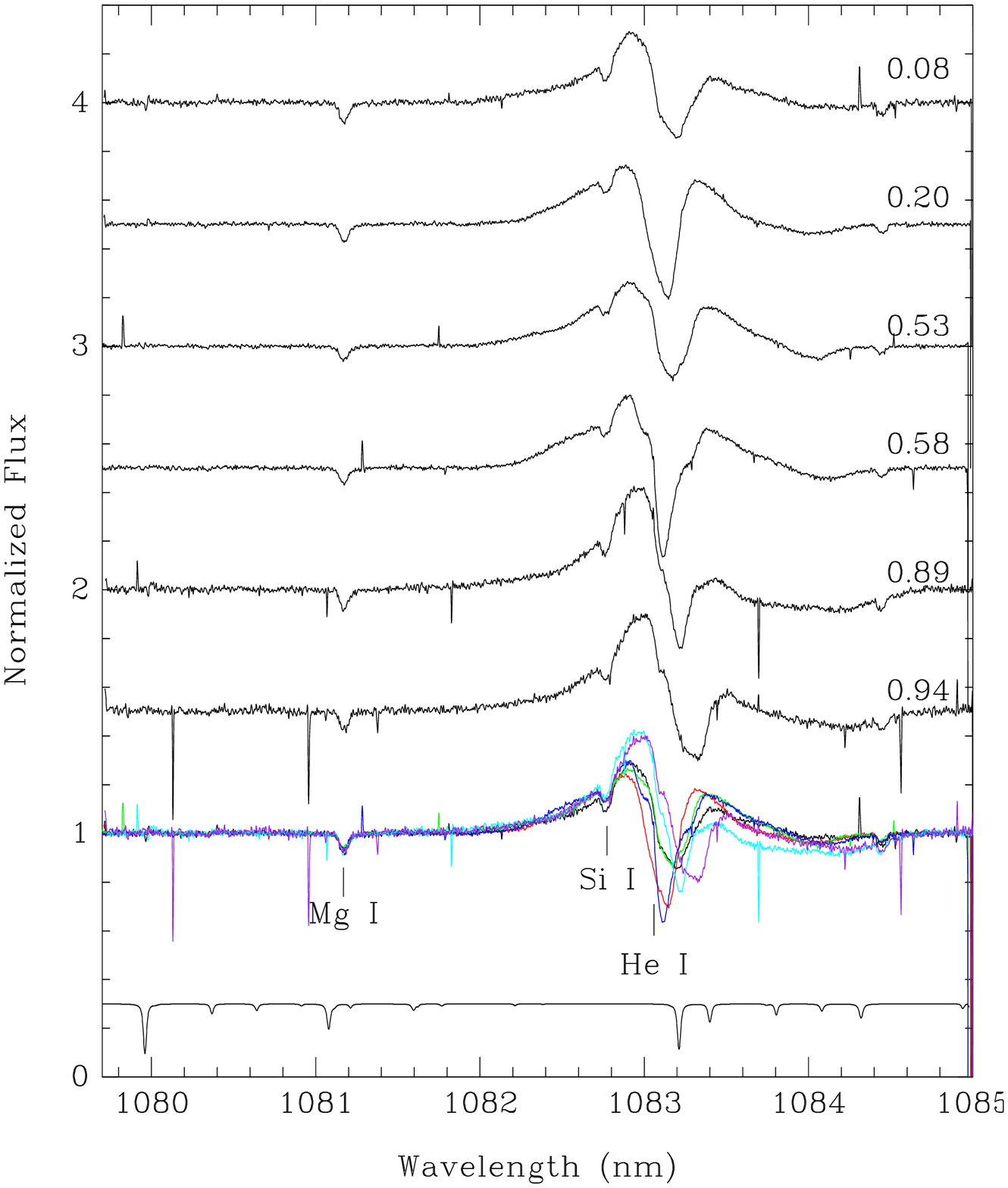}
\caption{
The variability of the \ion{He}{i} 1083.0\,nm line profile over the rotation period in the spectra of 
HD\,101412. The telluric 
spectrum presented at the bottom of the figure was removed from the observed spectrum.
The second spectrum from the bottom presents all six spectra overplotted.
}
\label{fig:hel}
\end{figure}

All lines identified in the CRIRES spectrum are presented in Table~\ref{tab:a3}.
The majority of the lines belong to the elements Mg and Si, followed by a few lines belonging to 
N, C, Fe, and Sr. Similar to the analysis at
optical wavelengths these elements, apart from N which is overabundant, are underabundant 
compared to solar abundances.
We find that iron is underabundant by $\sim$0.6\,dex in both optical and 
near-IR studies, and the magnetically sensitive \ion{Fe}{i} lines 1562.2\,nm, 1563.2\,nm,
and 1564.9\,nm appear rather weak in the CRIRES spectrum. No magnetic splitting is evident in these lines,
but weak profile variability of the stronger lines \ion{Fe}{i} 1562.2\,nm and 1563.2\,nm appears rather 
clear, confirming our previous results of Fe line profile variation in the visual wavelength range. 
In Fig.~\ref{fig:15640} we present the behaviour of the magnetically sensitive \ion{Fe}{i} lines at 
15621.7\,nm and 15631.9\,nm over the rotation period.
%Variable behaviour
Variability over the rotation period is also confirmed for the elements
Si and Mg, and is less established for N.
A few examples of the observed behaviour of the line profiles of \ion{Si}{i},
\ion{Mg}{i}, and \ion{N}{i} are presented in Figs.~\ref{fig:Sinorm}--\ref{fig:1076NI}.

\ion{He}{i} 1083.0\,nm is considered as a new 
diagnostic line to probe inflow (accretion) and outflow (winds) in the star-disk interaction region of 
accreting T\,Tauri and Herbig Ae/Be stars.
The uniqueness of this probe derives from 
the metastability of this transition and makes it a good indicator
of wind and funnel flow geometry (Edwards et al.\ \cite{Edwards2006}). Further, according to Edwards 
et al., the \ion{He}{i} 
line appears in emission for stronger mass accretion rates and in net absorption for low mass 
accretion rates.
Modeling of this line allowed Gregory et al.\ (2012, in preparation) for the first time to study the 
influence of field topologies on the star-disk interaction. Their models use magnetic fields 
with an observed degree
of complexity, as determined via field extrapolation from stellar magnetic maps. 
Also the recent work of  Adams \& Gregory (\cite{AdamsGregory2012}) shows
that high order field components may even
play a dominant role in the physics of the gas inflow as the accretion columns approach the star.
In Fig.~\ref{fig:hel} 
we present strong variability of this line over the rotation period.
The clear variations of the line profile
of the \ion{He}{i} line indicate that the magnetic field of this star is likely more 
complex than a dipole field. The work on the modeling of the field geometry on the surface of HD\,101412
using this line is currently ongoing
(Gregory et al.\ 2012, in preparation).

\begin{figure}
\centering
\includegraphics[angle=270,width=0.48\textwidth]{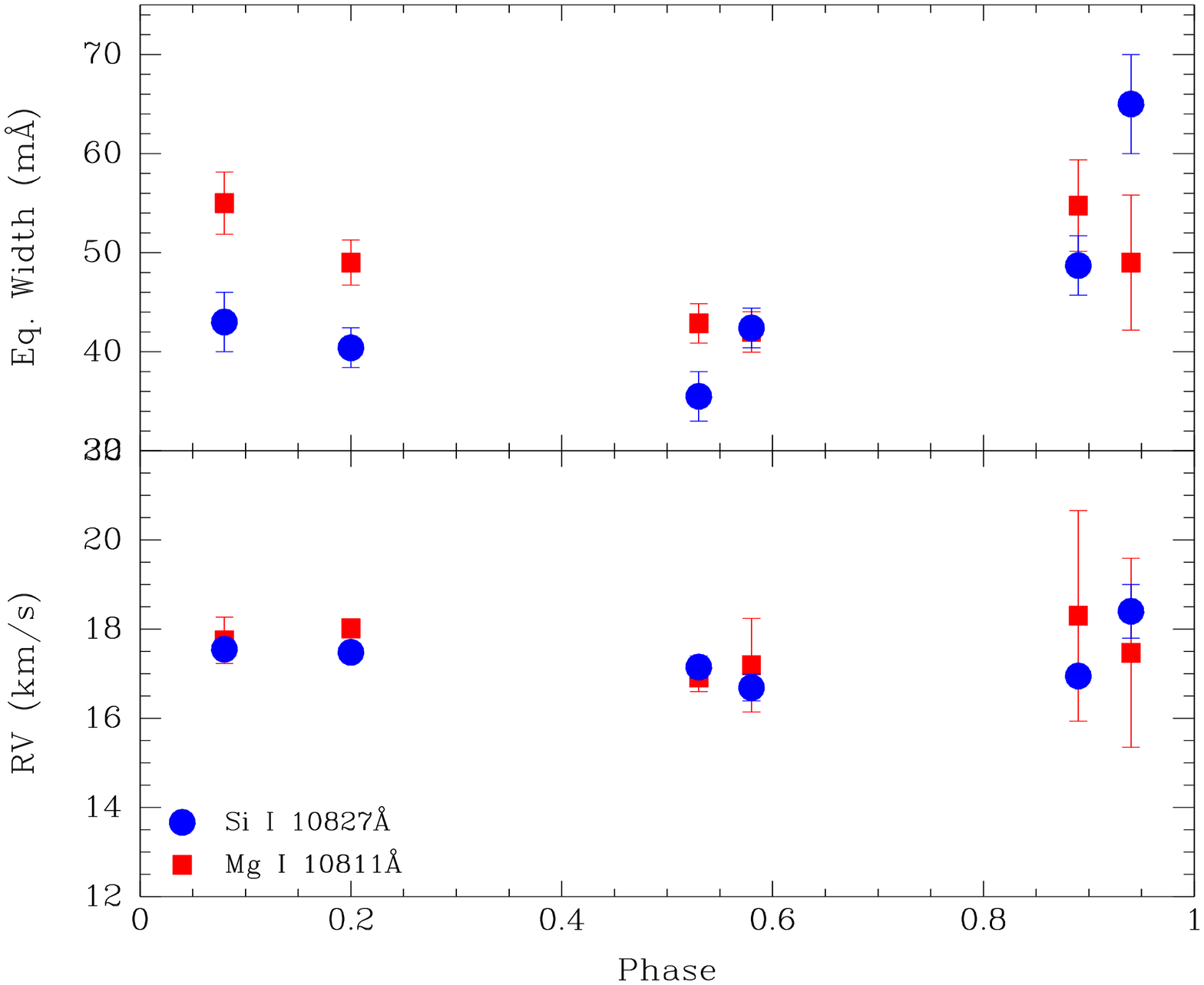}
\caption{
Variation of equivalent widths and radial velocities of the \ion{Mg}{i} 1081.1\,nm 
and \ion{Si}{i} 1082.7\,nm lines over the rotation period in the spectra of HD\,101412.
}
\label{fig:equiv}
\end{figure}

Measurements of equivalent widths and radial velocities of the most clean
spectral lines \ion{Mg}{i} 1081.1\,nm and \ion{Si}{i} 1082.7\,nm in the CRIRES spectrum 
confirm our previous finding on the existence of chemical
spots on the stellar surface (Hubrig et al.\ \cite{Hubrig2010}). The character of variations 
appears to be very similar to that found in optical
spectra where the equivalent width minimum is observed at the phase close to the positive extremum 
of the longitudinal magnetic field. 
In Fig.~\ref{fig:equiv} we present the variation of equivalent widths and radial velocities 
of the two lines \ion{Mg}{i} 1081.1\,nm and \ion{Si}{i} 1082.7\,nm.

No spectral lines belonging to exotic elements, such as the lanthanide rare earths, or heavier 
elements were identified in the spectrum of HD\,101412.

\subsection{51\,Oph (=HD\,158643)}

This target is a 0.7\,Myr old B9 Herbig star with $\teff=10,250\,K$ and $\logg=3.57$
(Montesinos et al.\ \cite{Montesinos2009} and references therein). According to Mora et al.\ 
(\cite{Mora2001}) it is one of the 
fastest rotating Herbig Ae/Be stars with $v \sin i=256$\,km\,s$^{-1}$. 
The huge rotational line broadening prevents the
reliable identification of spectral lines apart from the \ion{Mg}{i} line
at 1081.1\,nm, the \ion{He}{i} 1083.0\,nm line, and the Pa$\gamma$ line. We note, however, that 
we fitted these lines using $v \sin i$=100\,km\,s$^{-1}$. The large difference between this
value and the $v \sin i$ value determined by Mora et al.\
is probably due to our difficulties of putting
a proper continuum in the observed CRIRES spectrum covering the very short length of $\sim$5\,nm.

Interferometric observations with AMBER indicate that the system is observed almost edge-on 
with an inclination of the disk of $i= 82^{\circ}$ 
(Tatulli et al.\ \cite{Tatulli2008}).
This value is similar to the inclination $i=80\pm7^{\circ}$ of the disk of HD\,101412, 
which was derived from resolved 
observations of the disk using VLTI/MIDI (Fedele et al.\ \cite{Fedele2008}).
Using the value for the radius $R=5.6\,R_\odot$ from the work of Montesinos et al.\ and the
$v \sin i=256$\,km\,s$^{-1}$ from Mora et al.\ (\cite{Mora2001}), the expected rotation
period is of the order of 1.1\,d.

Not much is known about the presence of a magnetic field in this star.
FORS\,1 spectropolarimetric observations were carried out by Hubrig et al.
(\cite{Hubrig2009b}). The single measurement $\left<B_{\rm z}\right>$ = $32\pm 20$\,G  does not 
indicate the presence of a significant magnetic field during that observing epoch.

\begin{figure}
\centering
\includegraphics[angle=90,width=0.43\textwidth]{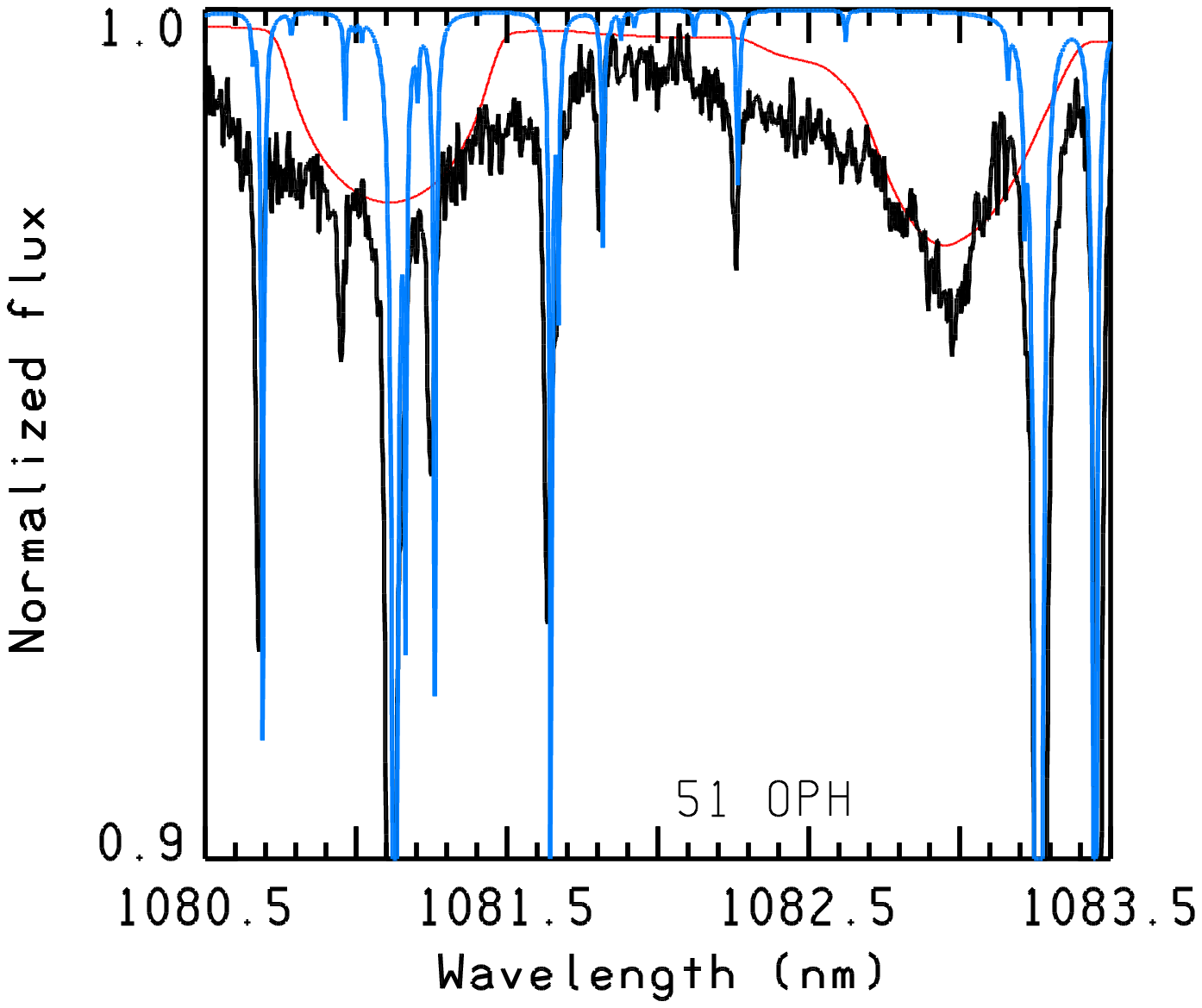}
\caption{ The \ion{He}{i} 1083.0\,nm and \ion{Mg}{i} 1081.1\,nm lines in the spectrum of the 
Herbig Be star 51\,Oph.}
\label{fig:51_he}
\end{figure}

\begin{figure}
\centering
\includegraphics[angle=90,width=0.43\textwidth]{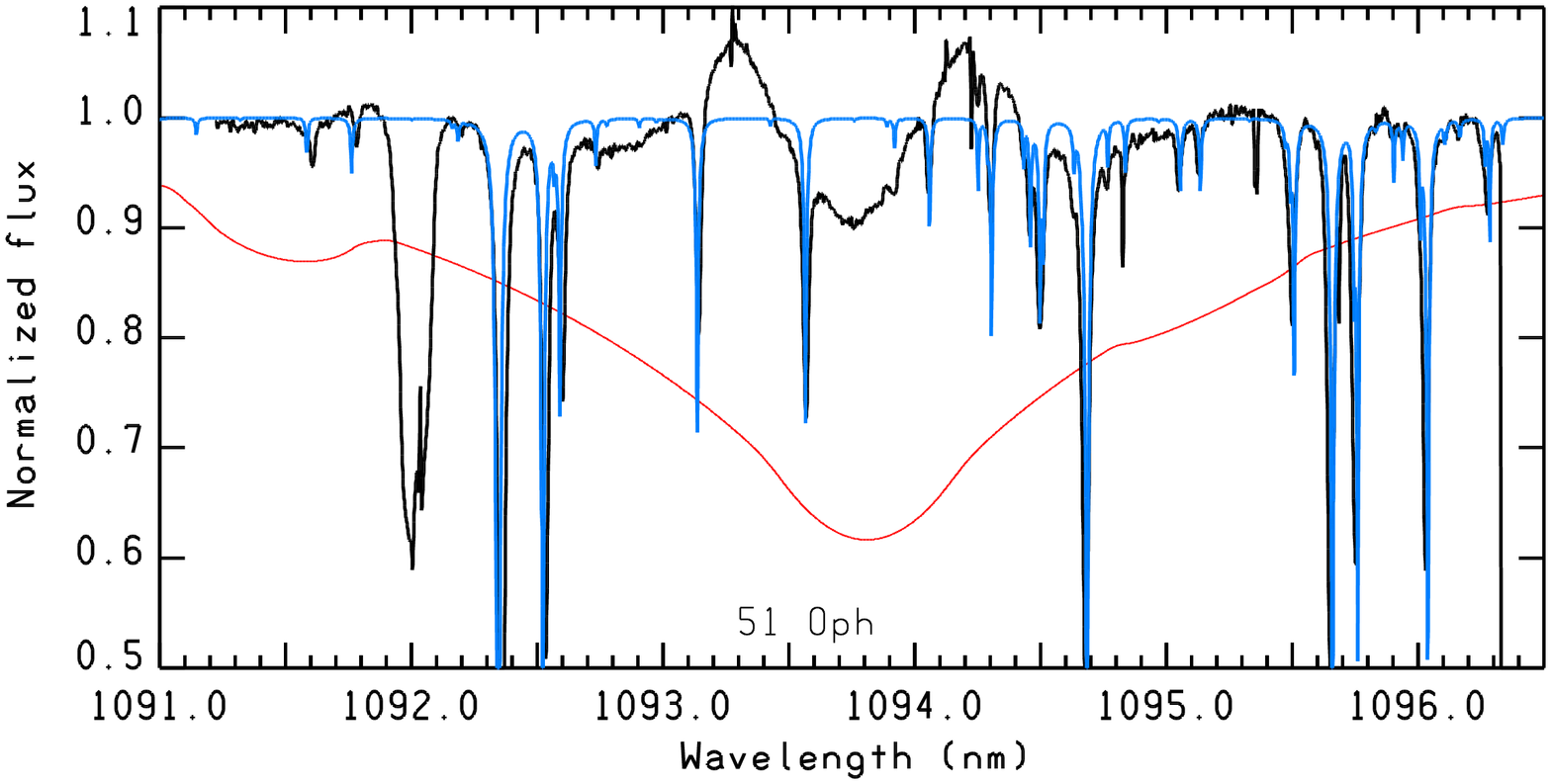}
\includegraphics[angle=90,width=0.43\textwidth]{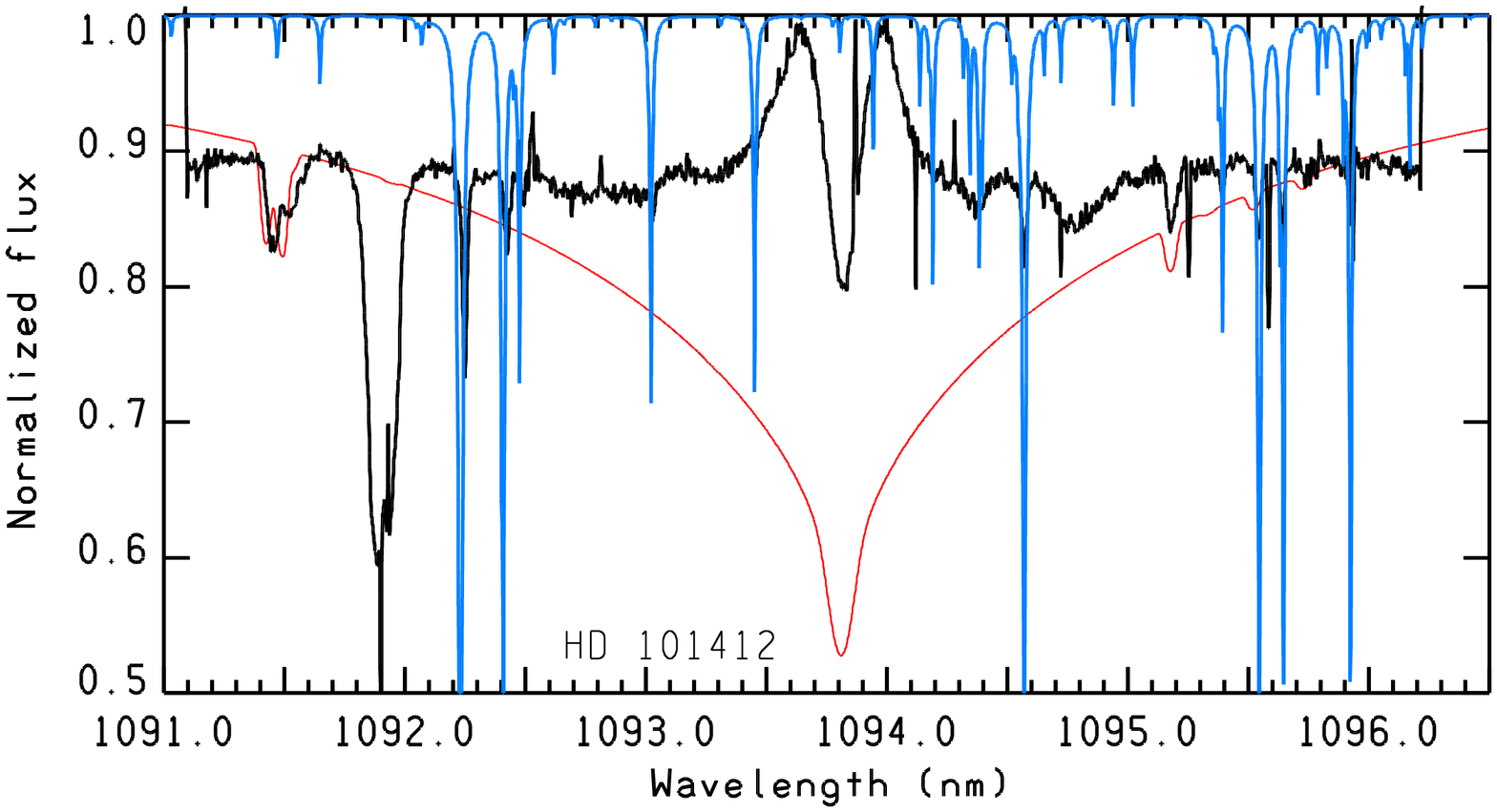}
\caption{The Pa$\gamma$ line in the spectra of the Herbig stars HD\,101412 and 51\,Oph.}
\label{fig:Paschen}
\end{figure}

In Figs.~\ref{fig:51_he} and \ref{fig:Paschen} we present the \ion{He}{i} 1083.0\,nm line, which appears in absorption,
and the Pa$\gamma$ line displaying two emission peaks, respectively. The shape of 
the emission profile resembles that of the Herbig Ae star HD\,101412 presented in the 
same figure, but is much broader. 

\section{Discussion}

In this work we use high quality near-IR spectra in a few spectral regions
to understand their diagnostic potential.
The presented line lists of identified and unidentified spectral features provide important 
line data for laboratory spectroscopists and for testing and improving models of stellar atmospheres.
The identification work clearly shows that the knowledge of the spectral features in near-IR 
wavelength regions is far from complete and that there is a strong need for additional atomic data, especially
for exotic elements, such as the lanthanide rare earths and heavy elements.
For example, the identification of rare earths lines is of special interest in the studies of roAp stars 
as the radial velocity oscillation amplitude for these lines is significantly higher than that for 
iron-peak elements (e.g. Gonz\'alez et al.,\cite{Gonz2008}).

Many lines in the spectra of the magnetic Ap stars appear magnetically split, but 
only for some of them are Land\'e factors known. The knowledge of magnetic sensitive lines
in the near-IR region will allow future studies to prove the presence of weak magnetic fields
in intermediate-mass stars not previously known to possess magnetic fields.

The content of the spectra of the studied Herbig Ae/Be stars is rather unexciting.
Variable behaviour of lines of the elements 
He, N, Mg, Si, and Fe in the spectra of the magnetic Herbig Ae star HD\,101412 over the rotation period 
confirm our previous finding of variability in the optical spectra.

Our work can be considered as a pilot project since the obtained CRIRES spectra cover a rather small 
part of the near-IR wavelength region. Clearly more complete
studies based on spectra obtained at all near-IR wavelengths from J to M band are needed 
to refine the analysis of intermediate-mass stars.

{
\acknowledgements
We thank E. Bi\'emont for providing us with the oscillator strengths for \ion{Ce}{iii} lines.
We would also like to thank the anonymous referee for his valuable comments.
Our research made use of Kurucz's atomic line lists and NIST and VALD databases. }

\appendix

\section{Tables presenting the line identifications for the two classical Ap stars $\gamma$\,Equ and 
HD\,154708, and the Herbig Ae star HD\,101412}

\begin{table*}[!hbp]
\caption[ ]{List of lines observed in the CRIRES spectrum of $\gamma$\,Equ.
The adopted parameters are $\teff=7700\,K$, $\logg=4.2$ (Ryabchikova et al.\ \cite{Ryabchikova1997}).
In the first column we present the observed wavelength followed by the central line depth
and the name of the element. Columns 4--8 list laboratory wavelength, oscillator strength,
excitation potentials for the lower and upper level of the transition and the source for the atomic data.
In the last column we make notes referring to the presence of blends, Zeeman splitting, or 
spectral artifacts.
The radial velocity shift between the observed wavelength positions of the spectral lines
and their laboratory wavelengths is 44\,km\,s$^{-1}$.
} 
\label{tab:a1}
\centering
\begin{tabular}{lllllrrll}
\hline\noalign{\smallskip}
\multicolumn{1}{c}{$\lambda$(obs)}&
\multicolumn{1}{c}{R$_{\rm c}$(obs)}&
\multicolumn{1}{l}{Element}&
\multicolumn{1}{c}{$\lambda$(lab)}&
\multicolumn{1}{l}{$\log\,gf$}&
\multicolumn{1}{c}{$\chi_{\rm low}$}&
\multicolumn{1}{c}{$\chi_{\rm up}$}&
\multicolumn{1}{l}{Source$^{a}$} &
\multicolumn{1}{l}{Notes}
\\
\multicolumn{1}{c}{[\AA{}]} &
 & &
\multicolumn{1}{c}{[\AA{}]} &
 &
\multicolumn{1}{c}{[cm$^{-1}$]} &
\multicolumn{1}{c}{[cm$^{-1}$]} \\
\hline\noalign{\smallskip}
 10729.4 & 0.80 & \ion{C}{i}  &  10729.530 & $-$0.420 & 60393.140 & 69710.660 &NIST4\\
 10733.7 & 0.95 & \ion{Ti}{ii} ?&  10733.742 & $+$0.530 & 62272.385 & 71586.249 & K10Ti2\\
 10737.6 & 0.90 & ?\\
 10741.9 & 0.95 & \ion{Si}{i} &  10741.729 & $-$1.100 & 53362.240 & 62669.179 & MB99\\
 10749.45 & 0.80 & \ion{Si}{i} &  10749.378 & $-$0.272 & 39760.285 & 49060.601 & NIST4 \\
 10749.8  & 0.83 & ?\\
 10754.25 & 0.92 & \ion{C}{i}  &  10753.975 & $-$1.606 & 60393.140 & 69689.480 & NIST4 \\
 10766.5 & 0.965& \ion{Si}{i} &  10766.449 & $-$1.419 & 54205.090 & 63490.660 & K07Si\\
 10768.5 & 0.98 & \ion{Al}{i} &  10768.345 & $-$1.560 & 32049.804 & 42233.722 & NIST4\\
 10769.0 & 0.975&  ? \\
 10770.1 & 0.97 & \ion{Si}{i} &  10770.134 & $-$1.253 & 53387.334 & 62669.727 & K07Si\\
 10770.6 & 0.96 & ? \\
 10771.2 & 0.97 & \ion{Fe}{i} &  10771.229 & $-$1.817 & 45061.326  & 54342.775 & K07 & \\
         &      & \ion{Ti}{ii} ?&  10771.294 & $+$0.399 & 62180.398  & 71461.791 & K10Ti2& \\
 10811.05& 0.77 & \ion{Mg}{i} &  10811.053 & $+$0.024 & 47957.045  & 57204.305 & NIST4 &\\
         &      & \ion{Mg}{i} &  10811.076 & $-$0.137 & 47957.027  & 57204.267 & NIST4 &\\
         &      & \ion{Mg}{i} &  10811.097 & $-$1.038 & 47957.045  & 57204.267 & NIST4 &\\
         &      & \ion{Mg}{i} &  10811.122 & $-$1.036 & 47957.027  & 57204.228 & NIST4 &\\
         &      & \ion{Mg}{i} &  10811.158 & $-$0.305 & 47957.058  & 57204.228 & NIST4 &\\
 10827.2 & 0.75 & \ion{Si}{i} &  10827.089 & $+$0.239 & 39955.053  & 49188.617 & NIST4 &\\
         &      & \ion{Ca}{i} &  10827.013 & $-$0.300 & 48568.950  & 39335.522 & K07Ca\\
 10829.5 & 0.97 & \ion{Ca}{i} &  10829.268 & $-$1.224 & 45050.419  & 35818.713 & K07Ca\\
         &      & \ion{Fe}{ii}&  10829.546 & $-$3.828 & 54275.637  & 45044.168 & K10\\
 10833.3 & 0.955& \ion{Ca}{i} &  10833.382 & $-$0.244 & 39335.322  & 48563.522 & K07Ca& blend telluric\\
 10839.25& 0.94 & \ion{Ca}{i} &  10838.970 & $+$0.238 & 39340.080  & 48563.522 & K07Ca& blend telluric\\
 10844.2 & 0.82 & \ion{Si}{i} &  10843.858 & $+$0.220 & 47284.061  & 56503.346 & NIST4\\
 10868.9 & 0.83 & \ion{Si}{i} &  10868.790 & $-$0.010 & 40993.775  & 59131.912 & MB99\\
 10869.65& 0.72 & \ion{Si}{i} &  10869.536 & $+$0.286 & 40991.884  & 50189.389 & NIST4\\
         &      & \ion{Ca}{i} &  10869.492 & $-$0.271 & 39340.080  & 48537.623 & K07Ca & blend telluric\\
 10871.6 & 0.96 & \ion{Fe}{ii}&  10871.601 & $-$3.034 & 45079.879  & 54275.637 & K10\\
         &      & \ion{Fe}{i} &  10871.491 & $-$0.490 & 50901.169  & 60097.020 & K07\\
 10872.6 & 0.95  & ?\\
 10882.8 & 0.89  & \ion{Si}{i}&  10882.809 & $-$0.646 & 48264.292  & 57450.580 & NIST4\\
 10885.35& 0.84  & \ion{Si}{i}&  10885.330 & $-$0.100 & 49850.830  & 59034.988 & MB99\\
 10891.8 & 0.98  & \ion{Al}{i}&  10891.720 & $-$1.110 & 32965.643  & 42144.402 & NIST4\\
 10894.8 & 0.98  & \ion{Si}{i}&  10894.792 & $-$1.680 & 49933.775  & 59109.959 & MB99\\
 10896.3 & 0.98  & \ion{Fe}{i}&  10896.299 & $-$2.692 & 24772.016  & 33946.931 & K07\\
 10905.7 & 0.97  & \ion{Cr}{i}&  10905.863 & $-$0.561 & 27728.812  & 36895.681 & K10Cr1\\
 10914.5 & 0.80  & \ion{Mg}{ii}& 10914.244 & $+$0.038 & 71490.190  & 80650.020 & NIST4\\
 10915.0 & 0.74  & \ion{Sr}{ii}& 10914.887 & $-$0.638 & 14555.900  & 23715.190 & WA \\
         &       & \ion{Mg}{ii}& 10915.284 & $-$0.918 & 71491.063  & 80650.020 & NIST4\\
 10938.2 & 0.55  & \ion{H}{i}  & 10938.086 & $+$0.002 &  9742.304  &109250.343 & NIST4\\
 10951.7 & 0.85  & \ion{Mg}{ii}& 10951.778 & $-$0.219 & 71491.063  & 80619.500 & NIST4\\
 10953.35& 0.92  & \ion{Mg}{i} & 10953.320 & $-$0.863 & 47841.119  & 56968.271 & NIST4\\
 10954.7 & 0.92  &      ?   \\
 10957.35& 0.86  & \ion{Mg}{i} & 10957.276 & $-$0.989 & 47844.414  & 56968.271 & NIST4\\
         &       & \ion{Mg}{i} & 10957.303 & $-$0.510 & 47844.414  & 56968.248 & NIST4 & blend telluric\\
 10979.3 & 0.86  & \ion{Si}{i} & 10979.308 & $-$0.562 & 39955.053  & 49060.601 & NIST4 & blend telluric\\
 10982.15& 0.88  & \ion{Si}{i} & 10982.058 & $-$0.270 & 49933.775  & 59037.043 & MB99\\
 10984.5 & 0.90  & \ion{Si}{i} & 10984.538 & $-$0.630 & 49933.775  & 59034.988 & MB99\\
 11013.6 & 0.92  & \ion{Si}{i} & 11013.703 & $-$1.310 & 50054.800  & 59131.912 & MB99\\
         &       & \ion{Fe}{i} & 11013.235 & $-$1.390 & 38678.036  & 47755.534 & K07\\
 11015.5 & 0.94  & \ion{Cr}{i} & 11015.679 & $-$0.420 & 27820.198  & 36895.681 & K10Cr1 & blend telluric\\
 11017.9 & 0.76  & \ion{Si}{i} & 11017.966 & $+$0.310 & 50054.800  & 59128.400 & MB99 & blend telluric\\
\\
 15505.5 & 0.84  &             &           &          &            &           &      &  artifact ?\\
 15532.0 & 0.95  &  \ion{Fe}{i}& 15531.752 & $-$0.236 & 45509.149  & 51945.814 & K07  &\\
         &       & \ion{Fe}{i} & 15531.805 & $-$0.281 & 50342.126  & 56778.769 & K07  &\\
 15532.5 & 0.945 & \ion{Si}{i} & 15532.449 & $-$1.397 & 54185.264  & 60621.640 & K07Si\\
\hline
\noalign{\smallskip}
\end{tabular}
\end{table*}

\addtocounter{table}{-1}
\begin{table*}[!hbp]
\caption[ ]{Continued.
} 
\centering
\begin{tabular}{lllllrrll}
\hline\noalign{\smallskip}
\multicolumn{1}{c}{$\lambda$(obs)}&
\multicolumn{1}{c}{R$_{\rm c}$(obs)}&
\multicolumn{1}{l}{Element}&
\multicolumn{1}{c}{$\lambda$(lab)}&
\multicolumn{1}{l}{$\log\,gf$}&
\multicolumn{1}{c}{$\chi_{\rm low}$}&
\multicolumn{1}{c}{$\chi_{\rm up}$}&
\multicolumn{1}{l}{Source$^{a}$} &
\multicolumn{1}{l}{Notes} \\
\multicolumn{1}{c}{[\AA{}]} &
 & &
\multicolumn{1}{c}{[\AA{}]} &
 &
\multicolumn{1}{c}{[cm$^{-1}$]} &
\multicolumn{1}{c}{[cm$^{-1}$]} \\
\hline\noalign{\smallskip}
 15533.5 & 0.948 &   ?   &\\
 15534.4 & 0.96 & \ion{Fe}{i} & 15534.245 & $-$0.384 & 45509.149  & 51944.781 & K07\\
 15535.3 & 0.965 & ? &\\
 15538.0 & 0.96 & \ion{Fe}{i} & 15537.697 & $-$0.031 & 50998.642  & 57432.844 & K07\\
 \\
 15541.5 & 0.97 & component &\\
 15542.2 & 0.96 & \ion{Fe}{i}&  15542.079& $-$0.336 & 45509.149 & 51941.537 & K07 & Zeeman splitting\\
 15542.9 & 0.96 & component\\
\\
 15550.0 & 0.97 & \ion{Fe}{i}& 15550.439 & $-$0.145 & 50998.642 & 57427.574 & K07& \\
 15551.5 & 0.95 & \ion{Fe}{i}& 15551.435 & $-$0.383 & 51192.270 & 57620.788 & K07&\\
 15555.4 & 0.96 & \ion{Ni}{i}& 15555.375 & $+$0.218 & 44262.599 & 50689.489 & K08\\
\\
         &      & \ion{H}{i} & 15556.467 & $-$1.166 &102823.904 & 109250.343& NIST4& lines hide the core \\
\\
15557.6 & 0.91 & component ?\\
         &       &\ion{Si}{i}& 15557.779 & $-$0.540 & 48102.323 & 54528.220 & NIST4& not observed !\\
 15558.25 & 0.91 & component ?\\
\\
 15561.1 &  0.96 & \ion{Fe}{i} &  15560.786 & $-$0.487 & 51207.995 & 57632.650 & K07\\
         &       & \ion{Si}{i} &  15561.251 & $-$1.456 & 56780.427 & 63204.890 & K07Si\\
\\
 15566.2 & 0.97  & component ?\\
 15566.85& 0.97  & \ion{Fe}{i} & 15566.727 & $-$0.428 & 51219.012 & 57641.215 & K07 & Zeeman splitting ?\\
 15567.5 & 0.975 & comonent ?\\
\\
15571.9   &0.985 & \ion{Fe}{i}&  15571.751 &  $-$1.162 & 50981.009 & 57401.140 & K07\\
          &      & \ion{Fe}{i}&  15571.121 & $-$1.414  &  47420.225 & 53840.616 & K07\\
\\
15611.0   & 0.97 & component ?\\
          &      & \ion{Fe}{i} & 15611.146 & $-$3.765 & 27543.001 & 33946.931 & K07 & Zeeman splitting ?\\
15611.6   & 0.97 & component ?\\
\\
15613.2   & 0.965 & component\\
15613.7   & 0.965 & \ion{Fe}{i}&  15613.628 &$-$0.517 & 51219.012 & 57621.924 & K07 & Zeeman splitting\\
15614.2   & 0.96  & component\\
\\
15615.4   & 0.98  & \ion{Fe}{i} & 15615.420 & $-$1.710 & 57029.610 & 50627.433& K07\\
15618.0   & 0.98  & \ion{Fe}{i} & 15617.703 & $-$0.783 & 57029.610 & 50628.369& K07& uncertain identification \\
\\
15621.15  & 0.93  & component\\
15621.8   & 0.91  & \ion{Fe}{i} &  15621.654 & $+$0.586 &44677.003 & 51076.625 & K07 & Zeeman splitting\\
15622.5   & 0.93  & components\\
\\
15631.25  & 0.948 & component\\
          &       & \ion{Fe}{i} & 15631.386 & $-$1.055 & 47377.952 & 53773.590 & K07\\
15632.05  & 0.926 & \ion{Fe}{i} & 15631.948 & $+$0.122 & 43163.323 & 49558.731 & K07 & Zeeman splitting\\
15632.8   & 0.944 & component \\
\\
15647.2   & 0.978 & component\\
15648.6   & 0.966 & \ion{Fe}{i} & 15648.510 & $-$0.596 & 43763.977 & 50152.616 & K07& Zeeman splitting \\
15650.15  & 0.976 & component\\
\\
15652.2   & 0.976 & component\\
15652.95  & 0.966 & \ion{Fe}{i}& 15652.874 & $-$0.170 & 50377.905 & 56764.763 & K07& Zeeman splitting\\
15653.7   & 0.966 & component\\
\\
15661.5   & 0.96  & component&   &           &          &           &           &     blend telluric\\
15662.05  & 0.934& \ion{Fe}{i} &  15662.016 & $+$0.382 & 47005.503 & 53388.633 & K07& Zeeman splitting\\
15662.75  & 0.956& component\\
\hline
\noalign{\smallskip}
\end{tabular}
\end{table*}

\addtocounter{table}{-1}
\begin{table*}[!hbp]
\caption[ ]{ Continued.
} 
\centering
\begin{tabular}{lllllrrll}
\hline\noalign{\smallskip}
\multicolumn{1}{c}{$\lambda$(obs)}&
\multicolumn{1}{c}{R$_{\rm c}$(obs)}&
\multicolumn{1}{l}{Element}&
\multicolumn{1}{c}{$\lambda$(lab)}&
\multicolumn{1}{l}{$\log\,gf$}&
\multicolumn{1}{c}{$\chi_{\rm low}$}&
\multicolumn{1}{c}{$\chi_{\rm up}$}&
\multicolumn{1}{l}{Source$^{a}$} &
\multicolumn{1}{l}{Notes} \\
\multicolumn{1}{c}{[\AA{}]} &
 & &
\multicolumn{1}{c}{[\AA{}]} &
 &
\multicolumn{1}{c}{[cm$^{-1}$]} &
\multicolumn{1}{c}{[cm$^{-1}$]} \\
\hline\noalign{\smallskip}
15664.7   & 0.98 & component&   &           &          &           &           &     blend telluric\\
15665.3   & 0.97 & \ion{Fe}{i}  &  15665.243 & $-$0.338 & 48221.321 & 54603.136 & K07& Zeeman splitting\\
15665.9   & 0.984& component\\

15674.8   & 0.96 & \ion{Si}{i} & 15674.653 & $-$1.150 & 56978.256 & 63356.24 & MB99\\
15677.3   & 0.95  & \ion{Fe}{i} & 15677.015 & $-$0.112 & 50377.905 & 56754.928 &K07\\
15716.0:  & 0.95  & \ion{Ce}{iii} & 15715.837 & $-$3.080 & 0.000 & 6361.000& BIE & on the wing of a telluric\\
          &        & \ion{Fe}{i}  & 15677.521 & $-$0.587 & 50377.905 & 56754.722 &K07\\
15723.65  & 0.936 & \ion{Fe}{i} &  15723.586 & $+$0.334 & 45333.872 & 51692.007 & K07\\
15740.8   & 0.83  & \ion{Mg}{i} &  15740.716 & $-$0.212 & 47841.119 & 54192.335 & NIST4&\\
15749.0   & 0.80  & \ion{Mg}{i} & 15748.886 & $-$0.338 & 47844.414 & 54192.335 & NIST4\\
          &       & \ion{Mg}{i} & 15748.988 & $+$0.140 & 47844.414 & 54192.294 & NIST4\\
15765.9   & 0.77  & \ion{Mg}{i} &  15765.645 & $-$1.514 & 47851.162 & 54192.335 & NIST4\\
          &       & \ion{Mg}{i} &  15765.747 & $-$0.337 & 47851.162 & 54192.294 & NIST4\\
          &       & \ion{Mg}{i} &  15765.842 & $+$0.411 & 47851.162 & 54192.256 & NIST4\\
15769.5   & 0.87  & \ion{Fe}{i} &  15769.423 & $+$0.700 & 44677.003 & 51016.657 & K07\\
\\
15773.7   & 0.932  & component\\
15774.0   & 0.932  & \ion{Fe}{i} &  15774.071 & $+$0.529 & 50807.994 & 57145.780 & K07 & Zeeman splitting\\
15774.3   & 0.928  & component\\
\\
15818.2   & 0.88   & \ion{Fe}{i} & 15818.142 & $+$0.576  & 45061.326 & 51381.454 & K07& blend telluric\\
          &        & \ion{Fe}{i} & 15819.134 & $+$0.009  & 50833.435 & 57153.167  & K07\\
15827.3   & 0.97   & \ion{Si}{i} & 15827.213 & $-$0.690  & 57198.027 & 63514.533 & MB99& blend telluric\\
15829.7   & 0.985  & ? \\
15833.68  & 0.88   & \ion{Si}{i}&  15833.602 &$-$0.450 & 50189.389& 56503.346 & MB99\\
15835.1   & 0.91   & \ion{Fe}{i}&  15835.167 &$+$0.738&  50833.435 &57146.768 & K07\\
15837.2   & 0.942   & \ion{Fe}{i}&  15837.079 &$-$0.993&   50901.169 & 57213.740 & K07\\
15837.9   & 0.94   &\ion{Fe}{i} & 15837.646 & $+$0.327 & 50833.435 & 57145.780 & K07\\
15840.35  & 0.945  & ?\\
15847.8   & 0.68   & \ion{Ce}{iii}& 15847.550 & $-$1.030 & 1528.320 & 7836.720 & BIE\\
15853.1   & 0.88   & \ion{C}{i}  & 15852.576 & $-$0.258 & 77679.820 & 83986.220 & NIST4\\
\hline
\noalign{\smallskip}
\end{tabular}
\tablefoot{
$^{a}$ BIE:~Biemont (2011, private communication),\\
       MB99:~Mel\'endez \& Barbuy (\cite{Mel1999}),\\
       NIST4:~http://www.nist.gov/pml/data/asd.cfm,\\
       K07Si:~http://kurucz.cfa.harvard.edu/atoms/1400/gf1400.pos,\\
       K07Ca:~http://kurucz.cfa.harvard.edu/atoms/2000/gf2000.pos,\\
       K07:~http://kurucz.cfa.harvard.edu/atoms/2600/gf2600.pos,\\
       K08:~http://kurucz.cfa.harvard.edu/atoms/2800/gf2800.pos,\\
       K10Ti2:~http://kurucz.cfa.harvard.edu/atoms/2201/gf2201.pos,\\
       K10:~http://kurucz.cfa.harvard.edu/atoms/2601/gf2601.pos,\\
       WA:~Warner (\cite{Warner1968})}.
\end{table*}

\begin{table*}[!hbp]
\caption[ ]{List of lines observed in the CRIRES spectrum of HD\,154708.
The adopted parameters are $\teff=6800\,K$, $\logg=4.11$ (Nesvacil et al.\ \cite{Nesvacil2008}).
The radial velocity shift between the observed wavelength positions of the spectral lines
and their laboratory wavelengths is 15\,km\,s$^{-1}$.}
\label{tab:a2}
\centering
\begin{tabular}{lllllrrll}
\hline\noalign{\smallskip}
\multicolumn{1}{c}{$\lambda$(obs)}&
\multicolumn{1}{c}{R$_{\rm c}$(obs)}&
\multicolumn{1}{l}{Element}&
\multicolumn{1}{c}{$\lambda$(lab)}&
\multicolumn{1}{l}{$\log\,gf$}&
\multicolumn{1}{c}{$\chi_{\rm low}$}&
\multicolumn{1}{c}{$\chi_{\rm up}$}&
\multicolumn{1}{l}{Source$^{a}$} &
\multicolumn{1}{l}{Notes} \\
\multicolumn{1}{c}{[\AA{}]} &
 & &
\multicolumn{1}{c}{[\AA{}]} &
 &
\multicolumn{1}{c}{[cm$^{-1}$]} &
\multicolumn{1}{c}{[cm$^{-1}$]} \\
\hline\noalign{\smallskip}
  10747.4   & 0.94 &  component\\
  10749.39 & 0.80 & \ion{Si}{i} & 10749.378 & $-$0.272 & 39760.285 & 49060.601 & NIST4\\
  10751.4   & 0.93 &component\\
  \\
  10769.7   & 0.98 &  ?\\
  10771.0   & 0.975&  ?  \\
  \\
  10809.8   & 0.92 &  component \\
  10811.1  & 0.84 & \ion{Mg}{i} & 10811.053  & $+$0.024 & 47957.045 & 57204.305 & NIST4\\
            &     & \ion{Mg}{i} & 10811.076  & $-$0.137 & 47957.027 & 57204.267 & NIST4\\
            &      &\ion{Mg}{i} & 10811.097  & $-$1.038 & 47957.045 & 57204.267 & NIST4\\
            &      &\ion{Mg}{i} & 10811.122  & $-$1.036 & 47957.027 & 57204.228 & NIST4\\
            &      &\ion{Mg}{i} & 10811.158  & $-$0.305 & 47957.058 & 57204.228 & NIST4\\
  10812.3   & 0.91 &  component \\
  10814.4   & 0.99 & ?\\
\\
  10825.2   & 0.92 &  component\\
  10827.1   & 0.75 & \ion{Si}{i} & 10827.089 & $+$0.239 & 39955.053 & 49188.617 & NIST4\\
  10829.2   & 0.93 & component\\
\\
  10834.3   & 0.96    & component &          &          &           &           &      & blend telluric\\
  10836.1   & 0.925&   \ion{Dy}{ii} & 10835.94 & $-$0.77 &16177.40  & 25343.42  & VALD \\
  10837.8   & 0.965     & component\\
\\
  10839.0  & 0.98  &  \ion{Ca}{i} & 10838.97 & $+$0.238 &39340.080 & 48563.522 & K07Ca\\
\\
  10842.5  & 0.92   &component \\
  10844.12 & 0.83   &\ion{Si}{i}& 10843.858 & $+$0.220& 47284.061 & 56503.346 & NIST4\\
  10845.7  & 0.92   & component\\
  10848.6  & 0.98&?\\
\\
  10867.3  & 0.94:&  guessed component &  &  & & & &blend\\
  10868.65 & 0.82 &\ion{Si}{i} & 10868.790 & $-$0.010 & 49933.775 & 59131.912 & MB99 & blend telluric\\
  10870.35 & 0.90 & guessed component\\
\\
  10868.0  & 0.88 &  guessed component & & & & & &blend\\
  10869.55 & 0.78 & \ion{Si}{i} & 10869.536 & $+$0.286 & 40991.884 & 50189.389& NIST4\\
  10870.8  & 0.90  & component\\
\\
  10882.02  &0.94 &   ? \\
  10883.5   &0.90 &   ? \\
  10884.85  &0.92 &   ? \\
\\
  10885.45 & 0.90  & \ion{Si}{i} & 1088.5333 & $-$0.100 & 49850.83 & 59034.988 & MB99\\
\\
  10914.3 &  0.85 & component\\
  10915.15 & 0.81 &\ion{Sr}{ii} & 10914.887 & $-$0.638 &14555.900 & 23715.190 & WA\\
  10916.1  & 0.88 & component \\
\\
  10936.7  & 0.74 & component ?\\
  10938.0  & 0.70 &  \ion{H}{i}& 10938.086 & $+$0.002 & 9742.304 & 109250.343 & NIST4\\
  10939.6  & 0.74 & component ?\\
\\
  10950.8  &0.96  &  ?\\
  10951.95 &0.94 & \ion{Mg}{ii}& 10951.778 ? & $-$0.219& 71491.063 & 80619.500 & NIST4\\
  10953.2  &0.935& \ion{Mg}{i}& 10953.320  ?  & $-$0.863 & 47841.119 & 56968.271 & NIST4\\
  10954.6  &0.94 & ?\\
  10954.85 &0.92 & ?\\
  10956.2  &0.91 & ?\\
  10958.0  &0.96 & ?\\
\hline
\noalign{\smallskip}
\end{tabular}
\end{table*}

\addtocounter{table}{-1}
\begin{table*}
\caption[ ]{Continued.}
\centering
\begin{tabular}{lllllrrll}
\hline\noalign{\smallskip}
\multicolumn{1}{c}{$\lambda$(obs)}&
\multicolumn{1}{c}{R$_{\rm c}$(obs)}&
\multicolumn{1}{l}{Element}&
\multicolumn{1}{c}{$\lambda$(lab)}&
\multicolumn{1}{l}{$\log\,gf$}&
\multicolumn{1}{c}{$\chi_{\rm low}$}&
\multicolumn{1}{c}{$\chi_{\rm up}$}&
\multicolumn{1}{l}{Source$^{a}$} &
\multicolumn{1}{l}{Notes} \\
\multicolumn{1}{c}{[\AA{}]} &
 & &
\multicolumn{1}{c}{[\AA{}]} &
 &
\multicolumn{1}{c}{[cm$^{-1}$]} &
\multicolumn{1}{c}{[cm$^{-1}$]} \\
\hline\noalign{\smallskip}
  10979.2  &0.84 & \ion{Si}{i}& 10979.308& $-$0.562& 39955.053 & 49060.601 & NIST4& blend telluric\\
  10980.6  &0.98 & ?\\
  10981.85 &0.90 & \ion{Si}{i} & 10982.058 & $-$0.270 &49933.775 & 59037.043 & MB99&\\
  10983.6  &0.90 &  ? \\
  10985.4  &0.96 &  ? \\
  11015.7 & 0.95 & \ion{Cr}{i} & 11015.679 & $-$0.42& 27820.198 & 36895.681 & K10Cr1\\
\\
  11016.2 &0.87 & component\\
  11017.9 &0.83 & \ion{Si}{i} & 11017.966 & $+$0.310 & 50054.800 & 59128.400 & MB99\\
  11019.3 &0.91 & component\\
\\ \\
  15528.0-15530. & 0.97  & ? & & & & & & broad \\
  15532.3        & 0.965 & ?\\
  15533.55       & 0.960 & ?\\
  15536.7        & 0.97  & ? & & & & & & broad\\
  15542.15       & 0.98  &  \ion{Fe}{i}& 15542.079& $-$0.336 & 45509.149 & 51941.537 & K07 \\
\\
  15556.0 & 0.96 & component ?\\
          &      &   \ion{H}{i} & 15556.467 & $-$1.166 &101823.904 & 109250.343& NIST4 & continuum\\
          &      &  \ion{Si}{i}  & 15557.779 & $-$0.540 &48102.323  & 54528.220 & NIST4   & not observed !\\
  15559.8 & 0.955 & component ?\\
\\
  15621.70  & 0.96 &\ion{Fe}{i}& 15621.654 & $+$0.586 & 44677.003 & 51076.625 & K07&\\
  15631.90  & 0.96 &\ion{Fe}{i}& 15631.948 & $+$0.122 & 43163.323 & 49558.731 & K07&\\
  15648.55  & 0.985&\ion{Fe}{i}& 15648.510 & $-$0.596 & 43763.977 & 50152.616 & K07&\\
  15653.00  & 0.99 &\ion{Fe}{i}& 15652.874 & $-$0.170 & 50377.905 & 56764.763 & K07&\\
  15661.95  & 0.98 &\ion{Fe}{i}& 15662.016 & $+$0.382 & 47005.503 & 53388.633 & K07&\\
  15665.25  & 0.985&\ion{Fe}{i}& 15665.243 & $-$0.338 & 48221.321 & 54603.136 & K07&\\
  15715.80  & 0.97 &\ion{Ce}{iii}& 15715.837&$-$3.080&0.000       & 8361.000& BIE& blend telluric\\
  15727.40  & 0.99 &\ion{C}{i}   &15727.376 & $-$0.682 & 77679.831&84036.327& NIST4&\\
\\
  15740.00  &0.98 & ? &&&&&& on telluric wing\\
 \\
  15746.0-15748.0 & 0.915 & component? &&&&& &broad\\
                  &       &\ion{Mg}{i}& 15748.886 &$-$0.338& 47844.414 & 54192.335& NIST4& not observed\\
                  &       &\ion{Mg}{i}& 14748.998 &$+$0.140& 47844.414 & 54192.294& NIST4& not observed\\
  15749.4 &0.90   &    component\\
  15750.3 &0.91   &    component\\
\\
  15762.90 &0.94& ?\\
\\
  15764.45 & 0.89&  component\\
  15766.0  &  ?  & \ion{Mg}{i} & 15765.645 & $-$1.514& 47851.162 & 54192.335 & NIST4& blend telluric\\
           &  ?  & \ion{Mg}{i} & 15765.747 & $-$0.337& 47851.162 & 54192.294 & NIST4&\\
           &  ?  & \ion{Mg}{i} & 15765.842 & $+$0.411& 47851.162 & 54192.256 & NIST4&\\
  15767.1  & 0.90&   component\\
\\
 15769.35  & 0.92 & \ion{Fe}{i} & 15769.423 & $+$0.700& 44677.003 & 51016.657 & K07&\\
 15770.50  & 0.93 & \ion{Fe}{i} & 15770.619 & $+$0.448& 50807.994 & 57147.167 & K07&blend telluric\\
 15822.70  & 0.98 & \ion{Fe}{i} & 15822.817 & $+$0.181& 45509.149 & 51827.410  & K07&\\
 15827.30  & 0.98 & \ion{Si}{i} & 15827.213 & $-$0.690& 57198.027 & 63514.533 &MB99&\\
\hline
\noalign{\smallskip}
\end{tabular}
\end{table*}

\addtocounter{table}{-1}
\begin{table*}
\caption[ ]{Continued.}
\centering
\begin{tabular}{lllllrrll}
\hline\noalign{\smallskip}
\multicolumn{1}{c}{$\lambda$(obs)}&
\multicolumn{1}{c}{R$_{\rm c}$(obs)}&
\multicolumn{1}{l}{Element}&
\multicolumn{1}{c}{$\lambda$(lab)}&
\multicolumn{1}{l}{$\log\,gf$}&
\multicolumn{1}{c}{$\chi_{\rm low}$}&
\multicolumn{1}{c}{$\chi_{\rm up}$}&
\multicolumn{1}{l}{Source$^{a}$} &
\multicolumn{1}{l}{Notes} \\
\multicolumn{1}{c}{[\AA{}]} &
 & &
\multicolumn{1}{c}{[\AA{}]} &
 &
\multicolumn{1}{c}{[cm$^{-1}$]} &
\multicolumn{1}{c}{[cm$^{-1}$]} \\
\hline\noalign{\smallskip}
  15830.0 &0.96& component ?\\
  15832.7 &0.945& component ?\\
          &     & \ion{Si}{i} & 15833.602 & $-$0.450& 50189.389 & 56503.346 & MB99& not observed !\\
  15834.4 &0.955 &   component ?\\
  15836.4 &0.945 &   component ?\\
\\
  15844.70 & 0.92 &    component\\
  15847.70 & 0.73 &   \ion{Ce}{iii} & 15847.550 & $-$1.030 & 1528.320 & 7836.720 & BIE &\\
  15850.80 & 0.88 &    component\\
\hline
\noalign{\smallskip}
\end{tabular}
\tablefoot{
$^{a}$BIE:~Bi\'emont (2011, private communication),\\
       MB99: Mel\'endez \& Barbuy (\cite{Mel1999}),\\
       NIST4:~http://www.nist.gov/pml/data/asd.cfm,\\
       K07:~http://kurucz.cfa.harvard.edu/atoms/2600/gf2600.pos,\\
       K07Si:~http://kurucz.cfa.harvard.edu/atoms/1400/gf1400.pos,\\
       VALD:~http://vald.astro.univie.ac.at/~vald/php/vald.php,\\
       WA:~Warner (\cite{Warner1968}).
}
\end{table*}

\begin{table*}[!hbp]
\caption[ ]{List of lines observed in the CRIRES spectrum of HD\,101412 at the phase 0.94. 
The adopted parameters are $\teff=8300\,K$, $\logg=3.8$ (Cowley et al.\ \cite{Cowley2010}).
The radial velocity shift between the observed wavelength positions of the spectral lines
and their laboratory wavelengths is $-$10\,km\,s$^{-1}$.} 
\label{tab:a3}
\centering
\begin{tabular}{lllllrrrll}
\hline\noalign{\smallskip}
\multicolumn{1}{c}{$\lambda$(obs)}&
\multicolumn{1}{c}{R$_{\rm c}$(obs)}&
\multicolumn{1}{l}{Element}&
\multicolumn{1}{c}{$\lambda$(lab)}&
\multicolumn{1}{l}{$\log\,gf$}&
\multicolumn{1}{c}{$\chi_{\rm low}$}&
\multicolumn{1}{c}{$\chi_{\rm up}$}&
\multicolumn{1}{l}{Source$^{a}$} &
\multicolumn{1}{l}{Notes} \\
\multicolumn{1}{c}{[\AA{}]} &
 & &
\multicolumn{1}{c}{[\AA{}]} &
 &
\multicolumn{1}{c}{[cm$^{-1}$]} &
\multicolumn{1}{c}{[cm$^{-1}$]} \\
\hline\noalign{\smallskip}
10749.4   & ? & \ion{Si}{i}&  10749.378 & $-$0.272 & 39760.285 & 49060.601 & NIST4& with artifact\\
10757.95  & 0.945 & \ion{N}{i}& 10757.887 & $-$0.389 & 95532.150 & 104825.110& NIST4&\\ 
10811.15  & 0.91  & \ion{Mg}{i} & 10811.053 & $+$0.024 & 47957.045 & 57204.305 & NIST4\\
          &       &  \ion{Mg}{i} & 10811.076 & $-$0.137 & 47957.027 & 57204.267 & NIST4\\
          &       & \ion{Mg}{i} & 10811.097 & $-$1.038 & 47957.045 & 57204.267 & NIST4\\
          &       &  \ion{Mg}{i} & 10811.122 & $-$1.036 & 47957.027 & 57204.228 & NIST4\\
          &       & \ion{Mg}{i} & 10811.158 & $-$0.305 & 47957.058 & 57204.228 & NIST4\\
10827.1   &       & \ion{Si}{i} & 10827.089 & $+$0.239 & 39955.053 & 49188.617 & NIST4 & on He I emission wing\\
10832.7   & 0.80  & \ion{He}{i} & 10829.091 & $-$0.745 & 159855.974 & 169087.831 & NIST4& shell line\\   
          &       & \ion{He}{i} & 10830.250 & $-$0.268 & 159855.974 & 169086.043 & NIST4& shell line\\   
          &       & \ion{He}{i} & 10830.340 & $-$0.046 & 159855.974 & 169086.766 & NIST4& shell line\\
10844.25   & 0.92  & \ion{Si}{i} & 10843.858 & $+$0.220 & 47284.061 & 56053.346 & NIST4\\
10869.6   & 0.92  & \ion{Si}{i} & 10869.536 & $+$0.286 & 40991.884 & 50189.389 & NIST4\\
10885.4   & 0.97  & \ion{Si}{i} & 10885.330 & $-$0.100 & 49850.830 & 59034.988 & MB99\\
10914.6   & 0.89  & \ion{Mg}{ii}& 10914.244 & $+$0.038 & 71490.190 & 80650.020 & NIST4\\
10915.3   & 0.90  & \ion{Sr}{ii}& 10914.887 & $-$0.638 & 14555.900 & 23715.190 & WA \\
          &       & \ion{Mg}{ii}& 10915.284 & $-$0.918 & 71491.063 & 80650.020 & NIST4\\
10938.6:  & 0.96: & \ion{H}{i}  & 10938.086 & $+$0.002 &  9742.304 &109250.343 & NIST4& emissions\\
10951.8   & 0.89  & \ion{Mg}{ii}& 10951.778 & $-$0.219 & 71491.063 & 80619.500 & NIST4\\
10979.4   & 0.96  & \ion{Si}{i} & 10979.308 & $-$0.562 & 39955.053 & 49060.601 & NIST4\\
10982.1   & 0.92: & \ion{Si}{i} & 10982.058 & $-$0.270 & 49933.775 & 59037.043 & MB99& blend telluric\\
11017.65  & 0.79: &  ?          &           &          &           &           &      & blend telluric\\  
11018.0   & 0.80: & \ion{Si}{i} & 11017.966 & $+$0.760 & 50054.800 & 59128.400 & K07Si& blend telluric\\
\\
 15727.55: & 0.96  & \ion{C}{i}  & 15727.376 & $-$0.682 & 77679.831 & 84036.327 & NIST4& blend telluric\\
 15740.7  & 0.975 & \ion{Mg}{i} & 15740.716 & $-$0.212 & 47841.119 & 54192.335 & NIST4\\
 15749.1  & 0.96  & \ion{Mg}{i} & 15748.886 & $-$0.338 & 47844.414 & 54192.335 & NIST4\\
          &       & \ion{Mg}{i} & 15748.998 & $+$0.140 & 47844.414 & 54192.294 & NIST4\\
 15765.8: &0.945  & \ion{Mg}{i} & 15765.645 & $-$1.514 & 47851.162 & 54192.335 & NIST4\\
          &       & \ion{Mg}{i} & 15765.747 & $-$0.337 & 47851.162 & 54192.294 & NIST4\\
          &       & \ion{Mg}{i} & 15765.842 & $+$0.411 & 47851.162 & 54192.256 & NIST4\\ 
15769.6:  &0.97:  &\ion{Fe}{i}  & 15769.423 & $+$0.700 & 44677.003 & 51016.657 & K07& blend telluric\\
15818.2:: &0.99   &\ion{Fe}{i}  & 15818.142 & $+$0.576 & 45061.326 & 51381.454 & K07\\
15822.9:: &0.99   &\ion{Fe}{i}  & 15822.817 & $+$0.181 & 45509.149 & 51827.410 & K07\\
15833.6:  &0.99   &\ion{Si}{i}  & 15833.602 & $-$0.450 & 50189.389 & 56503.346 & MB99\\
15852.85   &0.935  &\ion{C}{i}   & 15852.576 & $-$0.258 & 77679.820 & 83986.220 & NIST4\\
\hline
\noalign{\smallskip}
\end{tabular}
\tablefoot{
$^{a}$NIST4:~http://www.nist.gov/pml/data/asd.cfm,\\
       K07:~http://kurucz.cfa.harvard.edu/atoms/2600/gf2600.pos,\\
       MB99: Mel\'endez \& Barbuy (\cite{Mel1999}),\\
        %K07Si:~http://kurucz.cfa.harvard.edu/atoms/1400/gf1400.pos,\\
       WA:~Warner (\cite{Warner1968}).
}
\end{table*}

\end{document}